\documentclass[10pt]{article}
\usepackage{latexsym}
\usepackage{subfigure}
\usepackage{dcolumn}
\usepackage{bm}
\usepackage{ulem}
\usepackage{color}
\usepackage[colorlinks,linkcolor=magenta,anchorcolor=cyan,citecolor=blue]{hyperref}
\usepackage{amsmath,amssymb,multicol,fancyhdr,geometry}
\usepackage{graphicx}
\usepackage{cite}
\usepackage{float}
\usepackage{mathrsfs}
\usepackage{bm}
\newcommand{\eq}[1]{(\ref{#1})}

\newcommand{\fig}[1]{Fig.\,\ref{#1}}
\newcommand{\bra}[1]{|#1\rangle}
\def\be{\begin{equation}}
\def\ee{\end{equation}}
\def\ba{\begin{eqnarray}}
\def\ea{\end{eqnarray}}

\def\nn{\nonumber}
\def\lf{\left}
\def\rt{\right}

\def\nn{\nonumber}\def\lf{\left}\def\rt{\right}\def\q{\vartheta}
\def\w{\omega}  \def\y {\psi}   \def\p {\pi} \def\a {\alpha} \def\s
{\sigma} \def\d {\delta} \def\f {\phi} \def\g {\gamma}   \def\k {\kappa}  \def\z
{\zeta} \def\x {\xi}   
\def\m {\mu} \def\pd {\partial} \def \inf {\infty}  
\def\Q{\Theta} \def\W{\Omega} \def\Y {\Psi} 
   \def\D {\Delta}   \def\L {\Lambda}  

\def\.{\cdot}
\def\math {\mathcal}
\title{\textbf{Circuit complexity for free Fermion with a mass quench}}
\author{Jie Jiang\footnote{Email:jiejiang@mail.bnu.edu.cn}, Jieru Shan\footnote{Email:jierushan@mail.bnu.edu.cn}, and Jianzhi Yang\footnote{Email:201621140017@mail.bnu.edu.cn}\\
Department of Physics, Beijing Normal University,\\
Beijing 100875, China
}
\begin{document}
\maketitle
\begin{abstract}
By using a recent approach proposed by Hackl $et\, al.$ to evaluate the complexity of the free fermionic Gaussian state, we compute the complexity of the Dirac vacuum state as well as the excited state of the Fermi system with a mass quench. First of all, we review the counting method given by Hackl $et\, al.$, and demonstrate that the result can be adapted to all of the compact transformation group $G$. Then, we utilize this result to study the time evolution of the complexity of these states. We show that, for the rotational invariant reference state, the total complexity of the incoming vacuum state will saturate the value of the instantaneous vacuum state at the late time, with a typical timescale to achieve the final stable state. Moreover, we find that the complexity growth under the sudden quench is directly proportional to the mass difference, which shares similar behaviors with the holograph complexity growth rate in an AdS-Vaidya black hole with a shock wave, even though the dual boundary CFT is strongly coupled. Finally, we obtain some features of the excited state and the non-rotational reference state.

\end{abstract}
\section{Introduction}
Recently, the holography complexity has been used to understand how spacetime emerges from field theory degrees of freedom within the AdS/CFT correspondence\cite{A1, A2, A3, A4, A5, A6}. Two holographic proposals have been proposed by Susskind and others to describe the quantum complexity of state in boundary theory: the complexity=volume(CV) conjecture\cite{A1, A5} and the complexity=action(CA) conjecture\cite{A2, A6}. Meanwhile, there is a large number of papers developing and extending these ideas \cite{JZ,B1,B2,B3,B4,B5,B6,B7,B8,B9,B10,B11,B12,B13,B14,B15,B16}. The key to understand these conjectures is to understand what the complexity means in boundary CFT, quantum field theory and finite temperature field theory. Beyond obtaining a new perspective to the holographic complexity, developing the complexity of the field theory is an interesting research program in its own right. Recently, some researchers have provided a series of precise definitions of circuit complexity in quantum field theory\cite{C, C1, C2, C3, C4, C5, C6, C7}.

In this paper, we will follow the discussion in \cite{C}, where the circuit complexity of the quantum system derives from the computational complexity, which is the minimal number of the elementary gate necessary to implement. Nielsen and collaborators developed a geodesic approach to obtain the optimal circuit, which has been used to evaluate the complexity of the Gaussian state of the free scalar and fermion states\cite{C1, C2, C3}. These new approaches begin with identifying the discrete circuit for a target operator with a continuous curve $\g$ connecting identity and target operator in a certain Lie group $G$, where the Lie group $G$ is the transformation group of the target states. This operator can be regarded as the realization of the group on the target space, we can define its corresponding operator $U(g)$, such that
\ba
U(g)\bra{\Y_R}=\bra{\Y_T}.
\ea
Moreover, choosing the elementary gate is equivalent to defining a right-invariance cost function $F(\g,\dot{\g})$ in the Lie group $G$. Meanwhile, the optimal circuit can be obtained by minimizing the $cost$ which is defined by
\ba
\math{D}(\g)=\int^1_0dsF(\g(s),\dot{\g}(s)),
\ea
where the circuit $\g$ connects the reference state and target state.  And the cost function $F(\g,\dot{\g})$ is the local function along curve $\g$ and its tangent vector $\dot{\g}$. Next, we follow the discussion in \cite{C} to define the
cost function by the positive metric
$\langle\cdot,\cdot\rangle_g : T_gG\times T_gG\to \mathbb{R}$ on
the Lie group. $i.e.$,
\ba
F(\g,\dot{\g})\equiv\|\dot{\g}\|=\sqrt{\langle\dot{\g},\dot{\g}\rangle}_\g.
\ea
Then, minimizing this cost function is equal to obtaining the geodesic in the Lie Group $G$ equipped with this positive metric, and the complexity becomes the geodesic distance between identity and the target operator.

In this paper, our primary purpose is applying Nielsen's approach proposed by Hackl $et\, al.$\cite{C} to a dynamical, non-equilibrium Fermion system with a mass quench. In \cite{C}, the authors study the circuit complexity of the free Fermion system. As suggested by the holograph complexity, the dual boundary conformal field theory should be a strongly coupled system. Thus it is necessary for us to investigate the circuit complexity in a dynamical system.  It might be beneficial for the future study of the interacting theories to investigate the Fermion system with a mass quench. Meanwhile, it is hard to evaluate the circuit complexity for the thermodynamic system by Nielsen's approach. For the Fermion system with a mass quench, it has been shown in \cite{AC} that after the quench the entanglement entropy in momentum-space agrees with the prediction in a General Gibbs Ensemble, which means that studying the complexity following a mass quench might be a reflection in a thermodynamic system.

The structure of this paper is as follows: In section \ref{Circuitcomplexity}, we briefly review the approach proposed by Hackl $et\, al.$ and show that the results can also be adapted to any compact transformation group $G$, in which a bi-invariance metric can be defined. And then we apply this result to the Fermion system. In section \ref{Massquech}, we discuss the quantization of the Fermionic field with a mass quench. Then, in section \ref{Referenceandtarget}, we evaluate the covariant matrix of the reference state and the target state. In section \ref{Timeevolution}, we calculate the one-mode complexity and total complexity following a sudden quench and study their time evolution behaviours. Finally, in section \ref{conclusions}, we will summarize all the main results we draw out in previous sections.

\section{Circuit complexity}\label{Circuitcomplexity}
\subsection{Complexity for a general compact transformation group}
In this section, we will use the description of Nielsen's approach in the corresponding theoretic group language, where the circuit becomes a continuous curve under the transformation group $G$, $i.e.$
$\g:[0,1]\to G$, which satisfies the boundary condition
\ba\label{cd1}\begin{split}
\g(0)&=e\,,\\
U\lf(\g(1)\rt)\bra{\Y_R}&=\bra{\Y_T}\,.
\end{split}
\ea
Similar to Ref.\cite{C}, the cost function is defined as
\ba
F(\g(s))=\sqrt{\langle\dot{\g}(s),\dot{\g}(s)\rangle}_{\g(s)}\,,
\ea
where the metric $\langle\cdot,\cdot\rangle $ is positive and right-invariant. That is to say, it can be generated by the metric defined in the Lie Algebra $\math{G}=T_eG$, $i.e.\ \ \forall\, X,Y\in
T_g G$
\ba\label{met1}
\langle X,Y\rangle_g\equiv \langle R_g^{*}X,R_g^{*}Y\rangle_e.
\ea
In this paper, we only focus on the bi-invariant metric, $i.e.\ \ \forall\, X,Y\in T_g G$, we have
\ba\label{met2}
\langle L_g^{*}X, L_g^{*}Y\rangle_e=\langle X,Y\rangle_g \,.
\ea
\eq{met1} and \eq{met2} also imply $\forall\, g\in G$,
\ba
\langle Ad_g X, Ad_gY\rangle_e=\langle X,Y\rangle_e\,.
\ea
The complexity from a reference state $\bra{\Y_R}$ to a target state $\bra{\Y_T}$ can be obtained by minimizing the cost function, which means the complexity can be defined by
\ba
\math{C}(\bra{\Y_R}\to\bra{\Y_T})=\min_\g\int^1_0ds
F(\g(s),\dot{\g}(s))=\min_\g\int^1_0ds
\sqrt{\langle\dot{\g}(s),\dot{\g}(s)\rangle}_{\g(s)},
\ea
where the circuit $\g$ satisfies the boundary condition (\ref{cd1}). Evaluating this circuit complexity is equivalent to finding an
optimal circuit $\g$ which gives rise to the transformation
$U\lf(\g(1)\rt)\bra{\Y_R}=\bra{\Y_T}$. It's worth noting that if
there exists a stabilizer subgroup $H$ of the reference state
$\bra{\Y_R}$,  $i.e.$ $\forall\ h\in H$,
\ba
U(h)\bra{\Y_R}=\bra{\Y_R},
\ea
then there will be a lot of operators which can make the target state
invariant, $i.e.$
\ba
U(\g(1)h)\bra{\Y_R}=U(\g(1))\bra{\Y_R}=\bra{\Y_T}.
\ea
 Thus, obtaining the complexity is equivalent obtaining the optimal geodesic
 between $e$ and $g\in g_0 H$, in which
 $$U(g_0)\bra{\Y_R}=\bra{\Y_T}.$$ In order to obtain the complexity
 of the target state,  we first define the complexity of an operator
 $U(g)$
\ba\label{Cg}
\math{C}(g)=\min_\g\int^1_0ds
\sqrt{\langle\dot{\g}(s),\dot{\g}(s)\rangle}_{\g(s)},
\ea
where $\g$ satisfies
\ba
\g(0)=e,\ \ \ \ \g(1)=g.
\ea
That is to say, the complexity of this operator is the length of
the geodesic between $e$ and $g$. Thanks to the bi-invariance of
the metric, we can show the geodesic as an one-parameter subgroup
$e^{sA}$, which is generated by $A\in\math G$. According to \eq{met1}, one can obtain
\ba\begin{aligned}
{\langle\dot{\g}(s),\dot{\g}(s)\rangle}_{\g(s)}&=\langle
R_{\g(s)}^*\dot{\g}(s),R_{\g(s)}^*\dot{\g}(s)\rangle_e\\
&=\|A\|^2,
\end{aligned}\ea
where we have used
\ba\begin{aligned}
 R_{\g(s)}^*\dot{\g}(s)&=R_{\g(s)}^*\frac{d}{dt}\Big|_{t=s}\g(t)\\
 &=\frac{d}{ds}\Big|_{t=s}[\g(t)\g(-s)]\\
 &=\frac{d}{ds}\Big|_{s=0}{\g(s)}=A\,.
\end{aligned}\ea
From \eq{Cg}, the complexity of the operator $U(g)$ can be written
as
\ba
\math{C}(g)=\|A\|,
\ea
with $ g=e^A $. Thus, the complexity of the target state can be written as
\ba
\math{C}(\bra{\Y_R}\to\bra{\Y_T})=\min_{g\in g_0H}\math{C}(g).
\ea
To evaluate this complexity, we need to compare the value of the complexity for all of the operators $g\in g_0H$. Note that there exists a natural projection
\ba
\p:G\to G/H\,.
\ea
By this projection, we can define a vertical subspace $V_g,\ i.e.\
\forall\ g\in G$
\ba
V_g:=\ker(\p_{*g})\,.
\ea
Moreover, the corresponding horizontal subspace $H_g$ can be defined as
\ba
H_g:=\{X\in T_gG |\langle X,Y\rangle=0,\forall\ Y \in V_g\}\,.
\ea
Then, we have the decomposition
\ba
T_g G = V_g\oplus H_g\,.
\ea
Since the natural projection $\p_*:H_g\to T_{gH}(G/H)$ is an isomorphism, one can find that
$\forall \ \tilde{X}\in T_{gH}(G/H)$, there is only one
$\bar{X}_g\in H_g$, such that
\ba
\p_*\lf(\bar{X}_g\rt)=\tilde{X}\,,
\ea
where we call the vector $\bar{X}_g$ the horizontal lift vector of $\tilde{X}$ at point $g$. One can further prove  that $ \forall \,\tilde{X},\tilde{Y}\in T_{gH}(G/H)$, we have
\ba
\langle \bar{X}_g,\bar{Y}_g\rangle_g = \langle
\bar{X}_{gh},\bar{Y}_{gh}\rangle_{gh}\,,
\ea
Using these properties, we can define an induced metric in the  quotient group $G/H$
\ba
\langle \tilde{X},\tilde{Y}\rangle_{gH}:=\langle
\bar{X}_g,\bar{Y}_g\rangle_g\,.
\ea
One can verify that the curve $\p(\exp(tX))=\exp(tX)H$ is the
geodesic in the manifold $G/H$ equipped with this induced metric. And $\forall X_g\in T_g G$, we have
\ba
\langle X,X\rangle_{g}\geq\langle
\tilde{X},\tilde{X}\rangle_{gH}.
\ea

In order to obtain the complexity of the target state, we first focus
on a geodesic $\g$ between $e$ and $g\in g_0H$, then, we have
\ba\begin{aligned}
L(\g)&=\int_0^1ds\sqrt{\langle\dot{\g}(s),\dot{\g}(s)\rangle}_{\g(s)}\\
&\geq\int_0^1ds\sqrt{\langle\dot{\tilde{\g}}(s),\dot{\tilde{\g}}(s)\rangle}_{\tilde{\g}(s)}\\
&\geq\int_0^1ds\sqrt{\langle\dot{\bar{c}}(s),\dot{\bar{c}}(s)\rangle}_{\bar{c}(s)}\\
&=\int_0^1ds\sqrt{\langle\dot{c}(s),\dot{c}(s)\rangle}_{c(s)}\\
&= \|A\|,
\end{aligned}\ea
where $\tilde{\g}(s)=\p(\g(s))$, and the geodesic $c(t)=\exp{(t A)}$,
satisfying
\ba
c(1)\in g_0H,\ \  A\in H_e.
\ea
That is to say, the optimal geodesic between $e$ and $g\in g_0H$
can be generated by the  horizontal subspace $H_e$, and the
complexity of the target state can be expressed as
\ba\label{cplex}
\math{C}(\bra{\Y_R}\to\bra{\Y_T})= \|A\|,
\ea
where the Lie Algebra $A$ satisfies
\ba
A\in H_e\ \ \text{and} \ \ U(e^A)\bra{\Y_R}=\bra{\Y_T}.
\ea

\subsection{Fermionic Gaussian state}
In this section, we will focus on the complexity of the fermi
system, especially the fermionic Gaussian state. Ref.\cite{C}
illustrates that all information of the fermionic Gaussian
state can be reflected by their covariance matrix,
\ba
\langle\Y|\x^a\x^b\bra{\Y}=\langle\Y|\x^{[a}\x^{b]}\bra{\Y}+\langle\Y|\x^{(a}\x^{b)}\bra{\Y}=\frac{1}{2}\lf(i\W^{ab}+G^{ab}\rt),
\ea
where $\x^a\equiv(q_1,\cdots,q_N,p_1,\cdots,p_N)$ is the Majorana
modes of the $N$ degrees of freedom of fermions. $i.e.$, it should
satisfy
\ba
\{\x^a,\x^b\}=G^{ab}=\d^{ab}\,.
\ea
Thus the fermionic Gaussian state can be completely characterized by the antisymmetric part $\W^{ab}$. According to the discussion in the previous section, one can find that the complexity doesn't depend on the precise representation. So we will choose the description of the group given by their action on the covariant matrix $\W^{ab}$. For any Gaussian state $\bra{\tilde{\W}}$, there must exist a group of annihilation and creation operators $(\tilde{a}_i,\tilde{a}^{\dag}_i)$, such that
\ba
\tilde{a}_i\bra{\tilde{\W}}=0\,.
\ea
The corresponding Majorana modes can be constructed by
\ba
\tilde{q}_i=\frac{1}{\sqrt{2}}(\tilde{a}_i^{\dag}+\tilde{a_i})\,,\\
\tilde{p}_i=\frac{i}{\sqrt{2}}(\tilde{a}_i^{\dag}-\tilde{a_i})\,.
\ea
We denote $\tilde{\x}^a=(\tilde{q}_i,\tilde{p}_i)$, which
satisfies the anti-commutation relation
\ba\label{anc1}
\{\tilde{\x}^a,\tilde{\x}^b\}=\d^{ab}\,.
\ea
Moreover, one can also verify
\ba
\langle\tilde{\W}|[\tilde{\x}^a,\tilde{\x}^b]\bra{\tilde{\W}}=i
\W_0^{ab}= i
 \left(
\begin{array}{cc}
 0  & I  \\
 -I  & 0  \\
\end{array}
\right).
\ea
If we define a transformation $M$, such that
$\x^a=M^a_b\tilde{\x}^b$, then, from \eq{anc1}, we can verify that this transformation $M$ preserves the anti-commutation relation,
$i.e.$
\ba
(M M^T)^{ab}=\d^{ab}\,.
\ea
All of the transformations construct an $O(2N)$ group structure.
Hence, discussing the circuit complexity of the fermionic Gaussian
states means discussing the complexity of the Lie group $SO(2N)$.
To evaluate the complexity, next we need find the stabilizer subgroup $H$ of the reference state. First, we evaluate the covariance matrix of this state
\ba
i
\tilde{\W}^{ab}=\langle\tilde{\W}|[\x^a,\x^b]\bra{\tilde{\W}}=\langle\tilde{\W}|[{M^a}_c\tilde{\x}^c,{M^b}_d\tilde{\x}^d]\bra{\tilde{\W}}=i(M
\W_0 M^T)^{ab}\,.
\ea
Then, we have
\ba\label{WT}
\W_T=M_T\W_0{M_T}^T=M_T {M_R}^{T}\W_R M_R{M_T}^T=M \W_R M^T,
\ea
where we denote $M=M_T {M_R}^{T}\in SO(2N)$, and $\W_T,\W_R$ are
the covariance matrixes of the target state and reference state respectively. Moreover, the stabilizer subgroup $H$ should satisfy, $\forall\ \
M_s\in H$
\ba
M_s \W_R M_s^T=\W_R.
\ea
And the vertical subspace can be defined as
\ba
V_e=\{A_s\in\math{G}|\ [A_s,\W_R]=0\}\,.
\ea
Following the discussion in Ref.\cite{C}, we define the metric on the group $O(2N)$,
\ba
\langle A,B\rangle_e:= - Tr(A B),  \ \ \forall A,B\in T_e(O(2N))\,,
\ea
which is proportional to the Killing form. We can prove that this metric is bi-invariance. Considering any $B\in \math{G}$ which satisfies
\ba\label{tj1}
e^B\W_R=\W_R e^{-B}\,,
\ea
$i.e.$, $B\W_R=-\W_R B$, then, $\forall \ A_s\in H_e$, we have
\ba
\langle A_s,B\rangle_e= - \text{Tr}(A_s B) =  -\text{Tr}(A_s \W_R \W_R^{-1}B )= \text{Tr}(A_s B)=0\,,
\ea
which means $B\in H_e$. Moreover, one can verify that operator
$M=\sqrt{\W_T\W_R^{-1}}$ satisfies the condition (\ref{tj1}), so the
corresponding Lie Algebra $A=\log\sqrt{\W_T\W_R^{-1}}\in H_e $, and
\ba
M \W_R M^T=M^2 \W_R=\W_T\W_R^{-1}\W_R=\W_T,
\ea
From \eq{cplex}, we can obtain the complexity from the
reference state $\bra{\W_R}$ to the target state $\bra{\W_T}$
\ba\label{complexity}
\math{C}(\bra{\Y_R}\to\bra{\Y_T})= \|A\| =
\sqrt{-\text{Tr}\lf(\log\sqrt{\W_T\W_R^{-1}}\rt)^2}=\frac{\sqrt{\text{Tr}\lf((i
\log \D)^2\rt)}}{2}\,,
\ea
where
\ba
\D = \W_T {\W_R}^T
\ea
is called the relative covariance matrix.

\section{Mass quenches in free Fermionic field}\label{Massquech}
In this section, we make a quick review of the mass quenches for the Fermionic quantum field theory. The corresponding action can be written as
\ba
I=\int d^4 x\bar{\y}\left[i\g^\m\pd_\m-m(t)\right]\y\,,
\ea
where the mass profile $m(t)$ asymptotes to constants $m(-\inf)=m_\text{in}$ and $m(+\inf)=m_\text{out}$ at early and late times, respectively. This mode is equivalent to a standard quantum Dirac field of constant mass $m_0$ placed under a cosmological background and therefore can be understood from quantum field theory in curved spacetimes via intuition. And we can obtain the equation of motion
\ba\label{eom1}
\left[i\g^\m\pd_\m-m(t)\right]\y=0\,.
\ea
By virtue of the spatial symmetries, the solution can be written as
\ba
\y(\bm{x},t)=\int \frac{d^3\bm{k}}{(2\p)^{3/2}}\y(\bm{k},t)e^{i\bm{k}\.\bm{x}}.
\ea
Then, the equation becomes
\ba\label{eqyk}
\lf[i\g^0\pd_t-\bm{k}\.\bm{\g}-m(t)\rt]\y(\bm{k},t)=0\,.
\ea
To solve it, we consider a special formula
\ba\begin{aligned}
\y(\bm{k},t)&=[i\g^0\pd_t-\bm{k}\.\bm{\g}
+m(t)]f_{\bm{k}}(t)\,.
\end{aligned}\ea
Substituting it into \eq{eom1}, one can find that the function $f_{\bm{k}}(t)$ satisfies
\ba\label{eqf}
\ddot{f}_{\bm{k}}(t)+\lf[\bm{k}^2+m(t)^2-i\dot{m}(t)\g^0\rt]f_{\bm{k}}(t)=0\,.
\ea
The solution of this equation can be decomposed into
\ba
f_{\bm{k}}(t)=a_{\bm{k}}\f_{\bm{k}}(t)u_0+b_{-\bm{k}}^*\f^*_{\bm{k}}(t)v_0\,,
\ea
where the time independent four spinors $u_0, v_0$ are the eigenvectors of $\g^0$ with eigenvalues $\pm1$ respectively. Substituting it into \eq{eqf}, we have
\ba\label{fk}
\ddot{\f}_{\bm{k}}(t)+\lf[\bm{k}^2+m(t)^2-i\dot{m}(t)\rt]\f_{\bm{k}}(t)=0\,.
\ea
Here we focus on a special solution of $\f_{\bm{k}}$ which satisfies the asymptotic condition
\ba\label{lim}
\lim_{t\to-\inf}\f_{\bm{k}}(t)\sim e^{-i \w_{\text{in}}t}\,,
\ea
where we set $ \w_{\text{in/out}}=\sqrt{m_{\text{in/out}}^2+\bm{k}^2}\,.$ This solution describes the particle at the early time. Thus, we have
\ba
f_{\bm{k}}(t)=a_{\bm{k}}U_{\bm{k}}(t)e^{i\bm{k}\.\bm{x}}+b_{-\bm{k}}^*V_{-\bm{k}}(t)e^{i\bm{k}\.\bm{x}}\,,
\ea
where
\ba\label{UV}
U_{\bm{k}}(t)&=&[i\pd_t-\bm{\g}\.\bm{k}
+m(t)]\lf(\f_{\bm{k}}u_0\rt)\,,\\
V_{\bm{k}}(t)&=&[-i\pd_t+\bm{\g}\.\bm{k}
+m(t)]\lf(\f_{\bm{k}}^*v_0\rt)\,,
\ea
and we have used the relation $\f_{-\bm{k}}(t)=\f_{\bm{k}}(t)$. To obtain the explicit expression of this solution, we use the Weyl representation of the Dirac matrices, in which we define
\ba
\g^{\m}=
 \left(
\begin{array}{cc}
 0  & \s^{\m}  \\
 \bar{\s}^{\m}  & 0  \\
\end{array}
\right)
\ea
and $\s^\m\equiv(I,\bm{\s})$, $\bar{\s}^\m\equiv(I,-\bm{\s})$. Then, we have
\ba
u_0=
 \left(
\begin{array}{cc}
 \x \\
 \x
\end{array}
\right),
\ \ \ \ \
v_0=
 \left(
\begin{array}{c}
  \z \\
  -\z
\end{array}
\right)\,,
\ea
for any two-component spinors $\x$ and $\z$. And \eq{UV} can be reexpressed as
\ba\label{UV1}
U_{\bm{k}}^s(t)&=&\f_{\bm{k}}(t)
 \left(
\begin{array}{cc}
 \left[\w_{\bm{k}}(t)+m(t)-\bm{\s}\.\bm{k}\rt]\x \\
 \lf[\w_{\bm{k}}(t)+m(t)+\bm{\s}\.\bm{k}\rt]\x
\end{array}
\right)\,,\\
V_{\bm{k}}^s(t)&=&\f_{\bm{k}}^*(t)
 \left(
\begin{array}{cc}
 \left[\w_{\bm{k}}^*(t)+m(t)-\bm{\s}\.\bm{k}\rt]\zeta \\
 -\left[\w_{\bm{k}}^*(t)+m(t)+\bm{\s}\.\bm{k}\rt]\z
\end{array}
\right)\,,
\ea
where we set $\w_{\bm{k}}(t)=i\dot{\f}_{\bm{k}}/{\f}_{\bm{k}}$ as the frequency of this solution.

Next, we would like to find the explicit formula of this two spinors $\x^s$ and $\z^s$. According to \eq{eqyk}, one can find that $\y^{\dag}(\bm{k},t)\y(\bm{k},t)$ is the conserve quantity for any solution $\y$. Since $U_{\bm{k}}(t)e^{i\bm{k}\.\bm{x}}$ and $V_{\bm{k}}(t)e^{-i\bm{k}\.\bm{x}}$ are solutions, we can conventionally choose $U_{\bm{k}}^{\dag}U_{\bm{k}}=V_{\bm{k}}^{\dag}V_{\bm{k}}=1$. Using the early time behavior of $\f_{\bm{k}}$ (\ref{lim}), one can further obtain
\ba
\z^{\dag}\z=\x^{\dag}\x=\frac{1}{4\w_{\text{in}}(\w_{\text{in}}+m_{\text{in}})}\,.
\ea
Combining the condition $U_{\bm{k}}^{\dag}U_{\bm{k}}=V_{\bm{k}}^{\dag}V_{\bm{k}}=1$, it gives rise to the relation
\ba
|\f_{\bm{k}}(t)|\lf[\bm{k}^2+(m+\w_{\bm{k}})(m+\w_{\bm{k}}^*)\rt]=2
\w_{\text{in}}(\w_{\text{in}}+m_{\text{in}}).
\ea
For convenience, we choose two linearly independent spinors $\x$ and $\z$ as
\ba\label{xz}
\x^1=\z^1=\frac{1}{2\sqrt{\w_{\text{in}}(\w_{\text{in}}+m_{\text{in}})}}
 \left(
\begin{array}{c}
  1 \\
  0
\end{array}
\right)\,,\ \ \ \ \
\x^2=\z^2=\frac{1}{2\sqrt{\w_{\text{in}}(\w_{\text{in}}+m_{\text{in}})}}
 \left(
\begin{array}{c}
  0 \\
  1
\end{array}
\right)\,.
\ea
Then, the general solution of the field $\y$ can be written as
\ba
\y(x)=\int\frac{d^3k}{(2\p)^{3/2}}\sum_{s=1}^{2}\lf[a_{\bm{k}}^sU_{\bm{k}}^s(t)e^{i\bm{k}\.\bm{x}}
+b_{\bm{k}}^{s\dag}V_{\bm{k}}^s(t)e^{-i\bm{k}\.\bm{x}}\rt]\,.
\ea

Canonical quantization of the field $\y$, parameter $a_{\bm{k}}^s$ and $b_{\bm{k}}^s$ will become annihilation operators, and satisfy
\ba
\{{a}_{\bm{k}}^r,{a}_{\bm{k}'}^{s\dag}\}
=\{{b}_{\bm{k}}^r,{b}_{\bm{k}'}^{s\dag}\}
=\d_{rs}\d^{(3)}(\bm{k}-\bm{k}')\,.
\ea
Note that these operators correspond to the state at the early time.

\section{Reference and target states}\label{Referenceandtarget}
Next, we will evaluate the time evolution of the circuit complexities of some target states which are some particular states at the early time, such as the Dirac vacuum state or some excited states. Conventionally, we use the Heisenberg picture to define the quantum state. As suggested in \cite{C}, we will choose a reference state which is not only translationally invariant but also has no spatial entanglement. First, we focus on the reference state with the corresponding annihilation and creation operators which can be obtained by the expansion
\ba\label{yx}
\y(x)=\int\frac{d^3k}{(2\p)^{3/2}}
\sum_{s=1}^{2}\lf[\bar{a}_{\bm{k}}^s(t)u^s(M,\bm{q}) e^{i\bm{k}\.\bm{x}}
+\bar{b}^{s\dag}_{\bm{k}}(t)v^s(M,\bm{q})e^{-i\bm{k}\.\bm{x}}\rt]\,,
\ea
where
\ba
u^s(M,\bm{q})&=&
 \left(
\begin{array}{cc}
 \left(E_{\bm{q}}+M-\bm{\s}\.\bm{q}\rt)\x^s \\
 \lf(E_{\bm{q}}+M+\bm{\s}\.\bm{q}\rt)\x^s
\end{array}
\right)\,,\label{usvs1}\\
v^s(M,\bm{q})&=&
 \left(
\begin{array}{cc}
 \lf(E_{\bm{q}}+M+\bm{\s}\.\bm{q}\rt)\z^s \\
 -\lf(E_{\bm{q}}+M-\bm{\s}\.\bm{q}\rt)\z^s
\end{array}
\right)\,,\label{usvs2}
\ea
with $E_{\bm{q}}=\sqrt{\bm{q}^2+M^2}$, $\x^s=\z^s$ in \eq{xz}. Here $M$ is some mass scale and $\bm{q}$ is a special vector which is independent of the spacetime.  Then, the reference state at time $t$ can be defined as
\ba
\bar{a}_{\bm{k}}^s(t)\bra{\W_R(M,\bm{q},t)}=\bar{b}_{\bm{k}}^s(t)\bra{\W_R(M,\bm{q},t)}=0\,.
\ea
Note that this reference state is time-dependent. And one can verify that the correlation function of this reference state will vanish at each time $t$.
\subsection{Covariant matrix}
As shown in the previous section, the key to evaluating the circuit complexity is to obtain the covariant matrix of a Gaussian state. In order to construct a covariant matrix, we need to introduce an auxiliary state. For simplification, we choose $\lf(\y_i(\bm{k},t),\y_i^{\dag}(\bm{k},t),i=1,\cdots,4\rt)$ as its corresponding annihilation and creation operators, where we have
\ba
\{\y_i(\bm{k}),\y_j^{\dag}(\bm{k}')\}=\d_{ij}\d^{(3)}(\bm{k}-\bm{k}')\,,
\ea
with
\ba\label{yikt}
\y_i(\bm{k},t)=\int \frac{d^3\bm{x}}{(2\p)^{3/2}}\y_i(\bm{x},t)e^{-i \bm{k}\.\bm{x}}\,.
\ea
The Majorana modes can be defined as
\ba\label{QP}
Q_i(\bm{k},t)=\frac{1}{\sqrt{2}}\lf(\y_i^{\dag}(\bm{k},t)+\y_i(\bm{k},t)\rt)\,,\ \ \ \ \ P_i(\bm{k},t)=\frac{i}{\sqrt{2}}\lf(\y_i^{\dag}(\bm{k},t)-\y_i(\bm{k},t)\rt)\,.
\ea
According to the notation of section 3, we denote $\tilde{\x}^a(\bm{k},t)=\lf(Q_i(\bm{k},t),P_i(\bm{k},t)\rt)$. To obtain the covariant matrix, we only need obtain the transformation matrix $M$, such that $\x^a=M^a{}_b\tilde{\x}^b$, where $\x^a$ is the Majorana modes of the corresponding Gaussian state. Then, the covariant matrix can be written as $\W=M\W_0M^T$.
\subsubsection{Reference state}
Here, we calculate the covariant matrix of the reference state $\bra{\W_R(M,\bm{q},t)}$. The corresponding Majorana modes of this reference state can be defined as
\ba
\bar{q}^{s}_{\bm{k}}(t)&=&\frac{1}{\sqrt{2}}(\bar{a}^{s\dag}_{\bm{k}}(t)+\bar{a}^{s}_{\bm{k}}(t))\,,\ \ \ \ \ \ \bar{p}^{s}_{\bm{k}}(t)=\frac{i}{\sqrt{2}}(\bar{a}^{s\dag}_{\bm{k}}(t)-\bar{a}^{s}_{\bm{k}}(t))\,,\\
\bar{q'}^{s}_{\bm{k}}(t)&=&\frac{1}{\sqrt{2}}(\bar{b}^{s\dag}_{-\bm{k}}(t)+\bar{b}^{s}_{-\bm{k}}(t))\,,\ \ \ \bar{p}'^{s}_{\bm{k}}(t)=\frac{i}{\sqrt{2}}(\bar{b}^{s\dag}_{-\bm{k}}(t)-\bar{b}^{s}_{-\bm{k}}(t))\,,
\ea
and assemble these modes as $\bar{\x}^a(\bm{k},t)=\lf(\bar{q}^{s}_{\bm{k}}(t),\bar{q}'^{s}_{\bm{k}}(t),
\bar{p}^{s}_{\bm{k}}(t),\bar{p}'^{s}_{\bm{k}}(t)\rt)$.
For the annihilation and creation operators of this reference state, considering \eq{yx} and \eq{yikt}, one can obtain
\ba
\y_i(\bm{k},t)=\sum_{s=1}^{2}\lf[\bar{a}_{\bm{k}}^s(t)u_i^s(M,\bm{q})
+\bar{b}^{s\dag}_{-\bm{k}}(t)v_i^s(M,\bm{q})\rt]\,.
\ea
 By virtue of \eq{usvs1}, \eq{usvs2} and \eq{QP}, the covariant matrix of the reference state can be obtained by
\ba
\W_R=\oplus_{\bm{k}}\W_R(M,\bm{q}),
\ea
with
\ba\label{WR}
\W_R(M,\bm{q})=
\left(
\begin{array}{cccccccc}
 0 & \frac{q_y}{E_{\bm{q}} } & 0 & 0 & -\frac{q_z}{E_{\bm{q}} } & -\frac{q_x}{E_{\bm{q}} } & \frac{M}{E_{\bm{q}} } & 0 \\
 -\frac{q_y}{E_{\bm{q}} } & 0 & 0 & 0 & -\frac{q_x}{E_{\bm{q}} } & \frac{q_z}{E_{\bm{q}} } & 0 & \frac{M}{E_{\bm{q}} } \\
 0 & 0 & 0 & -\frac{q_y}{E_{\bm{q}} } & \frac{M}{E_{\bm{q}} } & 0 & \frac{q_z}{E_{\bm{q}} } & \frac{q_x}{E_{\bm{q}} } \\
 0 & 0 & \frac{q_y}{E_{\bm{q}} } & 0 & 0 & \frac{M}{E_{\bm{q}} } & \frac{q_x}{E_{\bm{q}} } & -\frac{q_z}{E_{\bm{q}} } \\
 \frac{q_z}{E_{\bm{q}} } & \frac{q_x}{E_{\bm{q}} } & -\frac{M}{E_{\bm{q}} } & 0 & 0 & \frac{q_y}{E_{\bm{q}} } & 0 & 0 \\
 \frac{q_x}{E_{\bm{q}} } & -\frac{q_z}{E_{\bm{q}} } & 0 & -\frac{M}{E_{\bm{q}} } & -\frac{q_y}{E_{\bm{q}} } & 0 & 0 & 0 \\
 -\frac{M}{E_{\bm{q}} } & 0 & -\frac{q_z}{E_{\bm{q}} } & -\frac{q_x}{E_{\bm{q}} } & 0 & 0 & 0 & -\frac{q_y}{E_{\bm{q}} } \\
 0 & -\frac{M}{E_{\bm{q}} } & -\frac{q_x}{E_{\bm{q}} } & \frac{q_z}{E_{\bm{q}} } & 0 & 0 & \frac{q_y}{E_{\bm{q}} } & 0 \\
\end{array}
\right)\,,
\ea
where we set the spatial vector $\bm{q}=\lf(q_x,q_y,q_z\rt)$. This one mode covariant matrix is independent of the momenta $\bm{k}$. And given parameters $M$ and $\bm{q}$  will give different reference states. What we should note is that, if we set $\bm{q}=0$, this state will be rotation invariant.

\subsubsection{Instantaneous vacuum state}
As a comparison, here we consider a series of special states, and each of them corresponds to a Dirac vacuum state $\bra{0(t)}$ at time $t$. The corresponding annihilation and creation operators can be obtained by the expansion
\ba\label{yxvac}
\y(x)=\int\frac{d^3k}{(2\p)^{3/2}}
\sum_{s=1}^{2}\lf[\breve{a}_{\bm{k}}^s(t)u^s(m(t),\bm{k}) e^{i\bm{k}\.\bm{x}}
+\breve{b}^{s\dag}_{\bm{k}}(t)v^s(m(t),\bm{k})e^{-i\bm{k}\.\bm{x}}\rt]\,.
\ea
By similar calculations, the covariant matrix of this state can be given by
\ba\label{Wvac}
\breve{\W}(t)=\oplus_{\bm{k}}\W_R(m(t),\bm{k})\,.
\ea

\subsubsection{Incoming vacuum state}
Here, we consider the target state which is the Dirac vacuum state $\bra{0}_{\text{in}}$ at early time. By virtue of \eq{UV1} and \eq{yikt}, one can find the relation between the target state and the auxiliary state, $i.e.$,
\ba\label{yx1}
\y_i(\bm{k},t)=\sum_{s=1}^{2}\lf[a_{\bm{k}}^sU_{\bm{k}}^s(t)
+b_{-\bm{k}}^{s\dag}V_{-\bm{k}}^s(t)\rt]\,.
\ea
Similarly, the Majorana modes of this target state can be defined as
\ba
q^{s}_{\bm{k}}&=&\frac{1}{\sqrt{2}}({a}^{s\dag}_{\bm{k}}+{a}^{s}_{\bm{k}})\,,\ \ \ \ \ \ p^{s}_{\bm{k}}=\frac{i}{\sqrt{2}}({a}^{s\dag}_{\bm{k}}-{a}^{s}_{\bm{k}})\,,\\
q'^{s}_{\bm{k}}&=&\frac{1}{\sqrt{2}}({b}^{s\dag}_{-\bm{k}}+b^{s}_{-\bm{k}})\,,\ \ \ p'^{s}_{\bm{k}}=\frac{i}{\sqrt{2}}(b^{s\dag}_{-\bm{k}}-b^{s}_{-\bm{k}})\,,
\ea
and assemble these modes as $\x^a(\bm{k},t)=\lf({q}^{s}_{\bm{k}},{q'}^{s}_{\bm{k}},
{p}^{s}_{\bm{k}},{p'}^{s}_{\bm{k}}\rt)$. With these in mind, the covariant matrix of the reference state can be obtained by
\ba
\W_T=\oplus_{\bm{k}}\W_T(\bm{k})
\ea
with
\ba\label{WTt}
\W_T(\bm{k})=\frac{|\f_{\bm{k}}|^2}{\w_{\text{in}}(\w_{\text{in}}+m_{\text{in}})}
\left(
\begin{array}{cccccccc}
 0 &k_y \a_{\bm{k}}  & -k_z {\d_{\bm{k}}} & -k_x {\d_{\bm{k}}} & -k_z \a_{\bm{k}}  & -k_x \a_{\bm{k}}  & {\k_{\bm{k}}} & -k_y {\d_{\bm{k}}} \\
 -k_y \a_{\bm{k}}  & 0 & -k_x {\d_{\bm{k}}} &k_z {\d_{\bm{k}}} & -k_x \a_{\bm{k}}  &k_z \a_{\bm{k}}  &k_y {\d_{\bm{k}}} & {\k_{\bm{k}}} \\
k_z {\d_{\bm{k}}} &k_x {\d_{\bm{k}}} & 0 & -k_y \a_{\bm{k}}  & {\k_{\bm{k}}} &k_y {\d_{\bm{k}}} &k_z \a_{\bm{k}}  &k_x \a_{\bm{k}}  \\
k_x {\d_{\bm{k}}} & -k_z {\d_{\bm{k}}} &k_y \a_{\bm{k}}  & 0 & -k_y {\d_{\bm{k}}} & {\k_{\bm{k}}} &k_x \a_{\bm{k}}  & -k_z \a_{\bm{k}}  \\
k_z \a_{\bm{k}}  &k_x \a_{\bm{k}}  & -{\k_{\bm{k}}} &k_y {\d_{\bm{k}}} & 0 &k_y \a_{\bm{k}}  & -k_z {\d_{\bm{k}}} & -k_x {\d_{\bm{k}}} \\
k_x \a_{\bm{k}}  & -k_z \a_{\bm{k}}  & -k_y {\d_{\bm{k}}} & -{\k_{\bm{k}}} & -k_y \a_{\bm{k}}  & 0 & -k_x {\d_{\bm{k}}} &k_z {\d_{\bm{k}}} \\
 -{\k_{\bm{k}}} & -k_y {\d_{\bm{k}}} & -k_z \a_{\bm{k}}  & -k_x \a_{\bm{k}}  &k_z {\d_{\bm{k}}} &k_x {\d_{\bm{k}}} & 0 & -k_y \a_{\bm{k}}  \\
k_y {\d_{\bm{k}}} & -{\k_{\bm{k}}} & -k_x \a_{\bm{k}}  &k_z \a_{\bm{k}}  &k_x {\d_{\bm{k}}} & -k_z {\d_{\bm{k}}} &k_y \a_{\bm{k}}  & 0 \\
\end{array}
\right)\,,
\ea
where we denote
\ba
\a_{\bm{k}}=m+\Re{\w_{\bm{k}}}\,,\ \ \ \d_{\bm{k}}=-\Im{\w_{\bm{k}}}\,,\ \ \
\k_{\bm{k}}=\frac{1}{2}(m+\w_{\bm{k}})(m+\w_{\bm{k}}^*)-\frac{1}{2}\bm{k}^2\,.
\ea
Here $\Re{\w_{\bm{k}}}$ and $\Im{\w_{\bm{k}}}$ denote the real and imaginary parts of $\w_{\bm{k}}$ separately.
At early time, we have
\ba
\a_{\bm{k}}\sim m+\w_{\text{in}}\,,\ \ \ \d_{\bm{k}}\sim 0\,,\ \ \
\k_{\bm{k}}\sim m_{\text{in}}(m_{\text{in}}+\w_{\text{in}})\,,
\ea
and the covariant matrix agree with the result of the instantaneous vacuum state found in the last section.

\subsubsection{Incoming excited state}\label{eset}
Note that for the fermion state, apart from vacuum state, the excited state is also a Gaussian state, which means one can apply this method to evaluate the complexity of an excited state. However, the particular state equipped with odd fermion number is on the disconnected component of the space for Gaussian states. Thus, we can only evaluate the complexity of Gaussian states with even fermion number. In this section, we consider a special excited state, which can be constructed by
\ba
\bra{\Y}=a_{\bm{k}}^{s\dag}b_{-\bm{k}}^{s\dag}\bra{0(t)}\,,
\ea
with arbitrary momenta $\bm{k}$. Here we choose spins aligned with the $z$-axis in the rest frame. Since
\ba
a_{\bm{k}}^{s\dag}\bra{\Y}=b_{-\bm{k}}^{s\dag}\bra{\Y}=0\,,
\ea
the corresponding annihilation operators can be given by $\lf(a_{\bm{k}}^{s\dag}, b_{-\bm{k}}^{s\dag},a_{\bm{k}'}^{r}, b_{-\bm{k}'}^{r}, (\bm{k}',r)\neq(\bm{k},s)\rt)$. Then, the Majorana modes of this target state can be defined as
\ba
q^{s}_{\bm{k}}&=&\frac{1}{\sqrt{2}}({a}^{s\dag}_{\bm{k}}+{a}^{s}_{\bm{k}})\,,\ \ \ \ \ \ p^{s}_{\bm{k}}=\frac{i}{\sqrt{2}}({a}^{s}_{\bm{k}}-{a}^{s\dag}_{\bm{k}})\,,\\
q'^{s}_{\bm{k}}&=&\frac{1}{\sqrt{2}}({b}^{s\dag}_{-\bm{k}}+b^{s}_{-\bm{k}})\,,\ \ \ p'^{s}_{\bm{k}}=\frac{i}{\sqrt{2}}(b^{s}_{-\bm{k}}-b^{s\dag}_{-\bm{k}})\,,
\ea
and
\ba
q^{r}_{\bm{k}'}&=&\frac{1}{\sqrt{2}}({a}^{r\dag}_{\bm{k}'}+{a}^{r}_{\bm{k}'})\,,\ \ \ \ \ \ p^{r}_{\bm{k}'}=\frac{i}{\sqrt{2}}({a}^{r\dag}_{\bm{k}'}-{a}^{r}_{\bm{k}'})\,,\\
q'^{r}_{\bm{k}'}&=&\frac{1}{\sqrt{2}}({b}^{r\dag}_{-\bm{k}'}+b^{r}_{-\bm{k}'})\,,\ \ \ p'^{r}_{\bm{k}'}=\frac{i}{\sqrt{2}}(b^{r\dag}_{-\bm{k}'}-b^{r}_{-\bm{k}'})\,,
\ea
when $(\bm{k}',r)\neq(\bm{k},s)$. Then, the covariant matrix of this state can be obtained by
\ba\label{WY}
\W_\Y=\lf[\oplus_{\bm{k}'\neq\bm{k}}\W_T(\bm{k}')\rt]\oplus \W_\Y(\bm{k})\,,
\ea
where
\ba
\W_\Y(\bm{k})=\frac{|\f_{\bm{k}}|^2}{\w_{\text{in}}(\w_{\text{in}}+m_{\text{in}})}
\left(
\begin{array}{cccccccc}
 0 & 0 & k_z \d_{\bm{k}}  & -2 k_y k_z & k_z \a_{\bm{k}}  & 0 & -\g_{\bm{k}}  & 2 k_x k_z \\
 0 & 0 & 2 k_y k_z & k_z \d_{\bm{k}}  & 0 & k_z \a_{\bm{k}}  & 2 k_x k_z & \g_{\bm{k}}  \\
 -k_z \d_{\bm{k}}  & -2 k_y k_z & 0 & 0 & -\g_{\bm{k}}  & 2 k_x k_z & -k_z \a_{\bm{k}}  & 0 \\
 2 k_y k_z & -k_z \d_{\bm{k}}  & 0 & 0 & 2 k_x k_z & \g_{\bm{k}}  & 0 & -k_z \a_{\bm{k}}  \\
 -k_z \a_{\bm{k}}  & 0 & \g_{\bm{k}}  & -2 k_x k_z & 0 & 0 & k_z \d_{\bm{k}}  & -2 k_y k_z \\
 0 & -k_z \a_{\bm{k}}  & -2 k_x k_z & -\g_{\bm{k}}  & 0 & 0 & 2 k_y k_z & k_z \d_{\bm{k}}  \\
 \g_{\bm{k}}  & -2 k_x k_z & k_z \a_{\bm{k}}  & 0 & -k_z \d_{\bm{k}}  & -2 k_y k_z & 0 & 0 \\
 -2 k_x k_z & -\g_{\bm{k}}  & 0 & k_z \a_{\bm{k}}  & 2 k_y k_z & -k_z \d_{\bm{k}}  & 0 & 0 \\
\end{array}
\right)
\ea
and
\ba
\g_{\bm{k}}=k^2-k_z^2+\k_{\bm{k}}\,.
\ea
One can see that this covariant matrix can be set as a perturbation of the vacuum state.

\section{Time evolution of the circuit complexity}\label{Timeevolution}
In this section, we consider a special case, which is the sudden quench limit of the general situation. Here, the mass profile $m(t)$ can be written as a step function, $i.e.$,
\ba
m(t)=m_{+}
+m_{-}\Q(t)\,,
\ea
where $\Q(t)$ is the step function, and
\ba
m_{\pm}=\frac{1}{2}(m_{\text{out}}\pm m_{\text{in}})\,.
\ea
Then, the equation of motion (\ref{fk}) becomes
\ba
\ddot{\f}_{\bm{k}}(t)+\lf[\bm{k}^2+m(t)^2-i m_{-}\d(t)\rt]\f_{\bm{k}}(t)=0\,,
\ea
which gives the  continuity conditions at $t=0$,
\ba
\f_{\bm{k}}(0^+)=\f_{\bm{k}}(0^-)=\f_{\bm{k}}(0)\,,\ \ \ \ \ \dot{\f}_{\bm{k}}(0^+)-\dot{\f}_{\bm{k}}(0^-)=i m_{-} \f_{\bm{k}}(0)\,.
\ea
By the asymptotic condition (\ref{lim}) at early time, one can further obtain $\f_{\bm{k}}(t)=e^{- i\w_{\text{in}}t}$ at $t<0$. Using the continuity conditions, one can obtain the solution
\ba\label{fgeq}
\f_{\bm{k}}(t)=\frac{2\w_{-}+m_{-}}{2\w_{\text{out}}}e^{i\w_{\text{out}}t}
+\frac{2\w_{+}-m_{-}}{2\w_{\text{out}}}e^{-i\w_{\text{out}}t}\,.
\ea
at $t\geq0$. Then, the frequency of this solution is given by
\ba\label{wk}
\w_{\bm{k}}(t)=\left\{
\begin{aligned}
  \w_{\text{in}} \ \ \ \ \ \ \ \ \ \ \ \ \ \ \ \ \ \ \ \ \ \ \ \ \ \ \ \ \ \ \ \ \ \ \ \ \ \ \ \ \ \ \ \ \ \ \ \ \ \ \ \ \ \ \ \ \ \ \ \ \ \ t<0 \\
  \frac{\omega _{\text{out}} \left[i \left(m_--\omega _{\text{in}}\right) \cos \left(\omega _{\text{out}}t\right)-\omega _{\text{out}} \sin \left( \omega _{\text{out}}t\right)\right]}{\left(m_--\omega _{\text{in}}\right) \sin \left(\omega _{\text{out}}t\right)-i \omega _{\text{out}} \cos \left(\omega _{\text{out}}t\right)}\ \ \ \ \ t>0
\end{aligned}
\right.\,.\ea

\subsection{Vacuum target state}
\subsubsection{Instantaneous vacuum state}
In this section, we consider the simple choice $\bm{q}=0$, which makes the reference state rotational invariant, $i.e.$, we set $\bm{q}=0$, and $E_{\bm{q}}=M$.
First, we consider the complexity from an instantaneous vacuum state at time $t$ to the rotational invariant reference state.  According to (\ref{WR}) and (\ref{Wvac}), the relative covariant matrix can be given by
\ba\breve{\D}(\bm{k},t)=
\left(
\begin{array}{cccccccc}
 \frac{m(t)}{\breve{\w}(t) } & 0 & -\frac{k_z}{\breve{\w}(t) } & -\frac{k_x}{\breve{\w}(t) } & 0 & 0 & 0 & -\frac{k_y}{\breve{\w}(t) } \\
 0 & \frac{m(t)}{\breve{\w}(t) } & -\frac{k_x}{\breve{\w}(t) } & \frac{k_z}{\breve{\w}(t) } & 0 & 0 & \frac{k_y}{\breve{\w}(t) } & 0 \\
 \frac{k_z}{\breve{\w}(t) } & \frac{k_x}{\breve{\w}(t) } & \frac{m(t)}{\breve{\w}(t) } & 0 & 0 & \frac{k_y}{\breve{\w}(t) } & 0 & 0 \\
 \frac{k_x}{\breve{\w}(t) } & -\frac{k_z}{\breve{\w}(t) } & 0 & \frac{m(t)}{\breve{\w}(t) } & -\frac{k_y}{\breve{\w}(t) } & 0 & 0 & 0 \\
 0 & 0 & 0 & \frac{k_y}{\breve{\w}(t) } & \frac{m(t)}{\breve{\w}(t) } & 0 & -\frac{k_z}{\breve{\w}(t) } & -\frac{k_x}{\breve{\w}(t) } \\
 0 & 0 & -\frac{k_y}{\breve{\w}(t) } & 0 & 0 & \frac{m(t)}{\breve{\w}(t) } & -\frac{k_x}{\breve{\w}(t) } & \frac{k_z}{\breve{\w}(t) } \\
 0 & -\frac{k_y}{\breve{\w}(t) } & 0 & 0 & \frac{k_z}{\breve{\w}(t) } & \frac{k_x}{\breve{\w}(t) } & \frac{m(t)}{\breve{\w}(t) } & 0 \\
 \frac{k_y}{\breve{\w}(t) } & 0 & 0 & 0 & \frac{k_x}{\breve{\w}(t) } & -\frac{k_z}{\breve{\w}(t) } & 0 & \frac{m(t)}{\breve{\w}(t) } \\
\end{array}
\right)\,,
\ea
where we denote $\breve{\w}(t)=\sqrt{k^2+m(t)^2}$. The corresponding eigenvalues appear with a multiplicity of four and are explicitly given by
\ba
\text{spec}(\D)=\frac{m(t)\pm i k}{\breve{\w}(t)}=e^{\pm2i\q}\,.
\ea
Then, the contribution to the complexity from each momentum and spin can be given by
\ba
\breve{Y}(\bm{k},s,t)=2\q=\tan^{-1}\lf(\frac{k}{m(t)}\rt)\,,
\ea
which completely agrees with the result found in \cite{C} for Dirac vacuum in the static Dirac system. The total complexity is then obtained by integrating over all momenta $\bm{k}$ and summing over the spins, $i.e.$,
\ba
\breve{\math{C}}_2(t)=\sqrt{V\int\frac{d^3\bm{k}}{(2\p)^3}\sum_{s}\breve{Y}(\bm{k},s,t)^2}.
\ea
For simplicity, now we consider the $\k=2$ definition of the complexity, $\math{C}_{\k=2}=\math{C}_2^2$, $i.e.$,
\ba
\breve{\math{C}}_{\k=2}(t)=V\int\frac{d^3\bm{k}}{(2\p)^3}\sum_{s}\breve{Y}(\bm{k},s,t)^2\,.
\ea
The one-mode complexty $\breve{Y}(\bm{k},s,t)\to\p/2$ at the limit of large momenta for any time. Whence, the total complexity  is UV divergent.
Choosing a hard cutoff $\L$ for the momentum integral, one can obtain the leading divergences of the total complexity
\ba\label{divC1}\begin{aligned}
\breve{\math{C}}_{\k=2}(t)&=\frac{V}{\p^2}\int^\L_{0} dkk^2\tan^{-1}\lf(\frac{k}{m(t)}\rt)\\
&\simeq\frac{V\L^3}{12}\lf(1-\frac{6m(t)}{\p \L}+\frac{12m^2(t)}{\p^2\L^2}\rt)\,,
\end{aligned}\ea
which shares the same divergence as the vacuum state at a static system.

\subsubsection{Incoming vacuum state}
In this section, we evaluate the complexity of the incoming vacuum state. Here we also choose the rotational invariant reference state. According to (\ref{WR}) and (\ref{WTt}), one can obtain the relative covariant matrix $\D$ between the target state and reference state,
\ba
\D=\oplus_{\bm{k}}\D(\bm{k})=\oplus_{\bm{k}}\lf[\W_T(\bm{k})\W_R^{-1}(M,\bm{0})\rt]\,,
\ea
with
\ba\label{DK1}
\D(\bm{k})=\frac{|\f_{\bm{k}}|^2}{\w_{\text{in}}(\w_{\text{in}}+m_{\text{in}})}
\left(
\begin{array}{cccccccc}
\k_{\bm{k}} & -k_y\d_{\bm{k}} & -k_z\a_{\bm{k}}  & -k_x\a_{\bm{k}}  & k_z\d_{\bm{k}} & k_x\d_{\bm{k}} & 0 & -k_y\a_{\bm{k}}  \\
 k_y\d_{\bm{k}} &\k_{\bm{k}} & -k_x\a_{\bm{k}}  & k_z\a_{\bm{k}}  & k_x\d_{\bm{k}} & -k_z\d_{\bm{k}} & k_y\a_{\bm{k}}  & 0 \\
 k_z\a_{\bm{k}}  & k_x\a_{\bm{k}}  &\k_{\bm{k}} & k_y\d_{\bm{k}} & 0 & k_y\a_{\bm{k}}  & -k_z\d_{\bm{k}} & -k_x\d_{\bm{k}} \\
 k_x\a_{\bm{k}}  & -k_z\a_{\bm{k}}  & -k_y\d_{\bm{k}} &\k_{\bm{k}} & -k_y\a_{\bm{k}}  & 0 & -k_x\d_{\bm{k}} & k_z\d_{\bm{k}} \\
 -k_z\d_{\bm{k}} & -k_x\d_{\bm{k}} & 0 & k_y\a_{\bm{k}}  &\k_{\bm{k}} & -k_y\d_{\bm{k}} & -k_z\a_{\bm{k}}  & -k_x\a_{\bm{k}}  \\
 -k_x\d_{\bm{k}} & k_z\d_{\bm{k}} & -k_y\a_{\bm{k}}  & 0 & k_y\d_{\bm{k}} &\k_{\bm{k}} & -k_x\a_{\bm{k}}  & k_z\a_{\bm{k}}  \\
 0 & -k_y\a_{\bm{k}}  & k_z\d_{\bm{k}} & k_x\d_{\bm{k}} & k_z\a_{\bm{k}}  & k_x\a_{\bm{k}}  &\k_{\bm{k}} & k_y\d_{\bm{k}} \\
 k_y\a_{\bm{k}}  & 0 & k_x\d_{\bm{k}} & -k_z\d_{\bm{k}} & k_x\a_{\bm{k}}  & -k_z\a_{\bm{k}}  & -k_y\d_{\bm{k}} &\k_{\bm{k}} \\
\end{array}
\right)\,.
\ea
The corresponding eigenvalues appear with a multiplicity of four and are explicitly given by
\ba
\text{spec}(\D)=\frac{|\f_{\bm{k}}|^2}{\w_{\text{in}}(\w_{\text{in}}+m_{\text{in}})}\lf(\k_{\bm{k}}\pm i |\bm{k}|\sqrt{\a_{\bm{k}}^2+\d_{\bm{k}}^2}\rt)=e^{\pm i2\q}\,.
\ea
Then, the contribution to the complexity from each momenta and spin can be given by
\ba
Y(\bm{k},s,t)=2\q=\tan^{-1}\lf(\frac{k\sqrt{\a_{\bm{k}}^2+\d_{\bm{k}}^2}}{\k_{\bm{k}}}\rt)=\tan^{-1}\lf(\frac{2k\sqrt{(m+\w_{\bm{k}})(m+\w_{\bm{k}}^*)}}{(m+\w_{\bm{k}})(m+\w_{\bm{k}}^*)-k^2}\rt)\,.
\ea
Substitute \eq{wk} into it, one can obtain
\ba
Y(\bm{k},s,t)=\left\{
\begin{aligned}
&\tan^{-1}\lf(\frac{k}{m_{\text{in}}}\rt)\ \ \ \ \ \ \ \ t<0\\
&\tan^{-1}\lf(\frac{2k\L_{\bm{k}}}{\L_{\bm{k}}^2-k^2}\rt)\ \ \ t>0
\end{aligned}
\right.\,,
\ea
with
\ba
\L_{\bm{k}}^2=\frac{C_1 \sin ^2\left(\omega _{\text{out}}t\right)+C_2\cos \left(2  \omega _{\text{out}}t\right)+C_3}{2 \left(m_--\omega _{\text{in}}\right){}^2 \sin ^2\left( \omega _{\text{out}}t\right)+2 \omega _{\text{out}}^2 \cos ^2\left( \omega _{\text{out}}t\right)}\,,
\ea
where
\ba
C_1&=&2m_{\text{out}}^2 \left(m_--\omega _{\text{in}}\right){}^2+2\omega _{\text{out}}^4\,,\\
C_2&=&\omega _{\text{out}}^2 \left(\left(m_--\omega _{\text{in}}\right){}^2+m_{\text{out}}^2\right)\,,\\
C_3&=&\omega _{\text{out}}^2 \left(-2 m_- \left(\omega _{\text{in}}+2 m_{\text{out}}\right)+4 \omega _{\text{in}} m_{\text{out}}+\omega _{\text{in}}^2+m_{\text{out}}^2+m_-^2\right)\,.
\ea
Note that in this case, the one-mode complexity jumps from the constant value at $t<0$ to an oscillatory behaviour at $t>0$ with the frequency $\w_{\text{out}}$, which has a similar behaviour as the scalar field case in \cite{AC2018}. It is worth noting that the frequency depends on the momentum $k$ (as shown in (c)), which will be integrated over the total complexity, $i.e.$, the total complexity will not have an exact frequency and amplitude at the late time. As illustrated in (a) and (b), the sign of the jump value for the one-mode complexity depends on the sign of $m_-$. For the mass-increasing quench $(m_->0)$, the complexity will shrink at $t=0$, while for the mass-decreasing quench $(m_-<0)$ it will grow, which implies that the total complexity will also possess the similar characteristics. In (d), we compare the one-mode complexity with the instantaneous vacuum state and the corresponding incoming vacuum state and find that when $t>0$, the equilibrium position for this state is the same as the instantaneous vacuum state.
\begin{figure}[H]
\centering
\includegraphics[width=6.5in,height=4in]{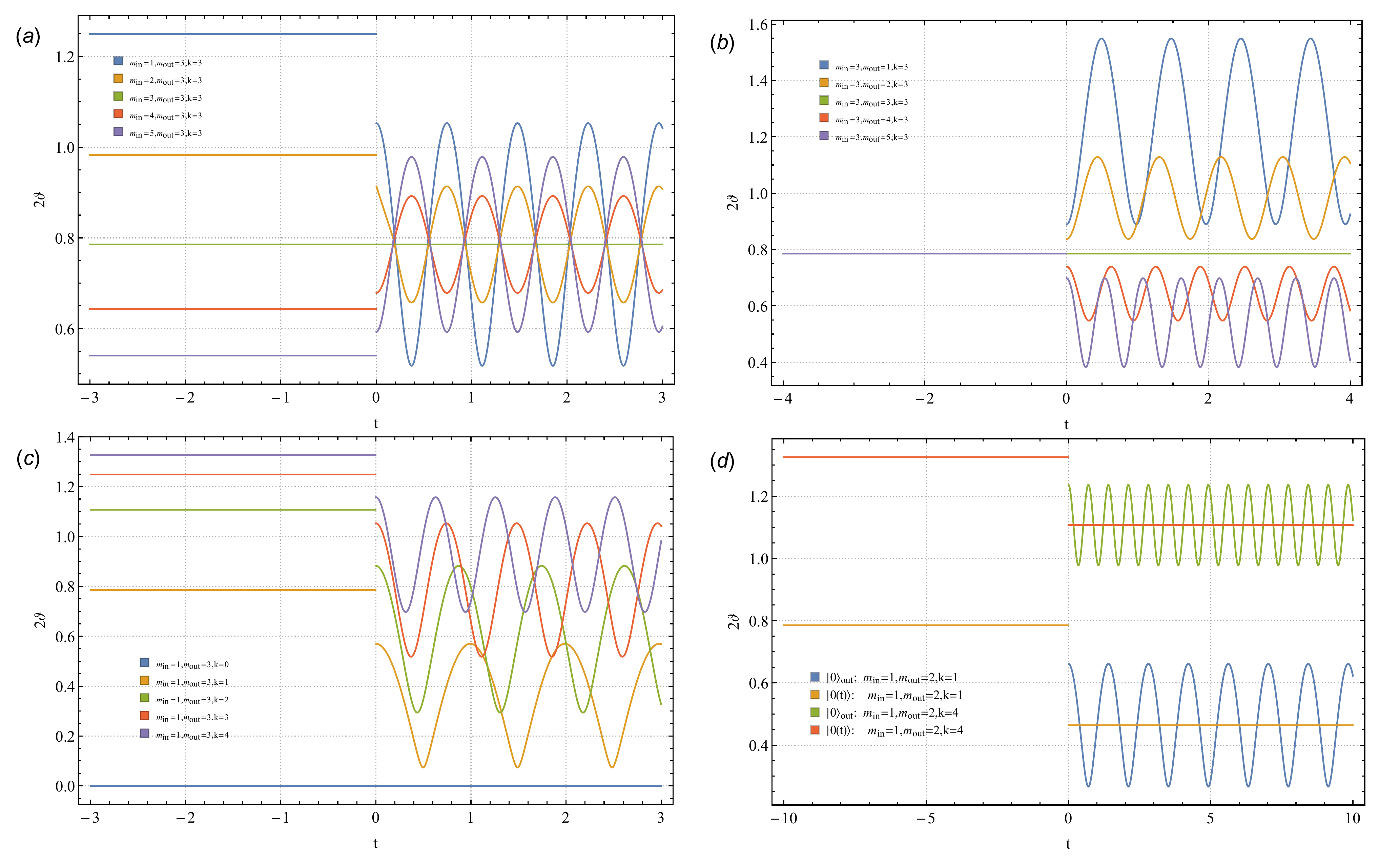}
\caption{Time evolution of the one-mode complexity. In figure (a), $m_{\text{out}}=3, k=3$ are fixed and $m_{\text{in}}$ is changing as shown in the figure. In (b), we fix $m_{\text{in}}=3, k=3$ and change $m_{\text{in}}$. In (c), we fix $m_{\text{in}}=1,m_{\text{out}}=3$ and vary the momenta $k$. And in (d), we compare the one-mode complexity of the incoming vacuum state with the instantaneous vacuum state.}
\label{fig1}
\end{figure}
Next, we turn to consider the total complexity. By virtue of \eq{wk}, one can find that $\w_{\bm{k}}\to \bm{k}$ at large momenta $\bm{k}$. That is to say, the one-mode complexity $Y(\bm{k},s,t)\to\p/2$ at the limit of large momenta. Hence, the total complexity is also UV divergent.
Choosing the same cutoff $\L$, The total complexity can be written as
\ba\begin{aligned}
\math{C}_{\k=2}(t)&=\frac{V}{\p^2}\int^\L_{0} dkk^2\tan^{-1}\lf(\frac{k\sqrt{\a_{\bm{k}}^2+\d_{\bm{k}}^2}}{\k_{\bm{k}}}\rt)\,.
\end{aligned}\ea
The relevant results are shown in \fig{fig2}. For numerical convenience here we fix $\L=100$. As stated above, the total complexity shares similar behaviours at $t=0$. Moreover, this figure also shows that the jump value of the full complexity is directly proportional to the mass difference $m_-=\d m$, same result as the difference between the late time complexity and the early time complexity. Considering the AdS/CFT correspondence, we have that the boundary QFT with a mass quench might dual to the AdS black hole with a shock wave, where the incoming vacuum state corresponds to the AdS vacuum, and the late time thermal state corresponds to the AdS-Vaidya hole. According to Ref.\cite{CMM}, the late time holograph complexity growth rate of the AdS-Vaidya black hole has the expression
\ba\label{Chol}
\dot{\math{C}}_\text{hol}\propto \d M\,,
\ea
where $\d M$ is the energy of the shock wave. To compare the holograph complexity with our circuit complexity, we define a relative complexity of this incoming vacuum state as
$\math{C}_\text{rlt}=\math{C}_\text{out}-\math{C}_\text{in}$. Then, we have
\ba\label{Crlt}
\math{C}_{\text{rlt}}\propto \d m\,.
\ea
It might be entirely different for these two results. However, note that the QFT in this paper is a free system, but the dual field should be a strongly coupled system. With a view to \eq{Chol} and \eq{Crlt}, we propose that the circuit complexity for a free system is dual to the complexity growth rate for a strongly coupled system, $i.e.$, we have
\ba\label{fs}
\math{C}_\text{free}\propto\dot{\math{C}}_{\text{strg}}\,.
\ea
By this conjecture and the CA conjecture, we can connect the free field complexity to the holograph complexity.

Furthermore, as shown in this figure, the amplitude has decreased significantly with the time evolution and finally shrink to zero. Then, from (b) of \fig{fig2}, one can find that the total complexity will saturate the result of the instantaneous vacuum state at the late time, and there exists a typical timescale to achieve the finial stable state. This result means that the total complexity will share some similar divergent behaviours with \eq{divC1}.
However, as shown in \fig{fig2}, By virtue of the finite amplitude at the finite time $t>0$, the amplitude of the total complexity will also diverge when $\L\to\inf$, which means that apart from the divergence \eq{divC1}, there also exist some divergent parts contributed by the amplitude. This result might imply that all of the Dirac vacuum states are analogous at the late time under the complexity perspective. Moreover, this might give the common feature of the Dirac vacuum although these vacuums are totally different under the mass quench.

\begin{figure}[H]
\begin{minipage}[b]{0.5\linewidth}
\centering
\includegraphics[width=3in,height=2in]{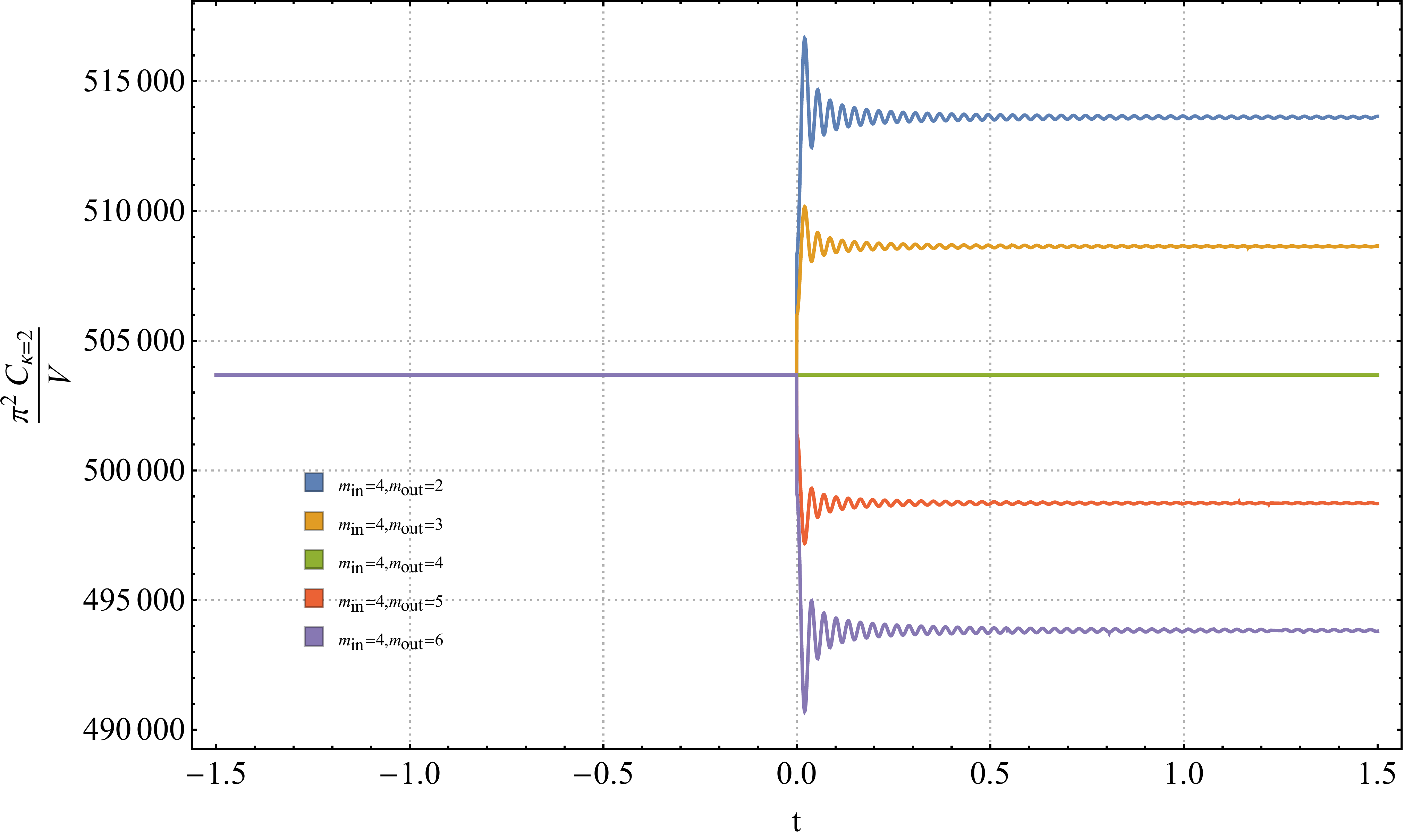}
\end{minipage}
\begin{minipage}[b]{.5\linewidth}
\includegraphics[width=3in,height=2in]{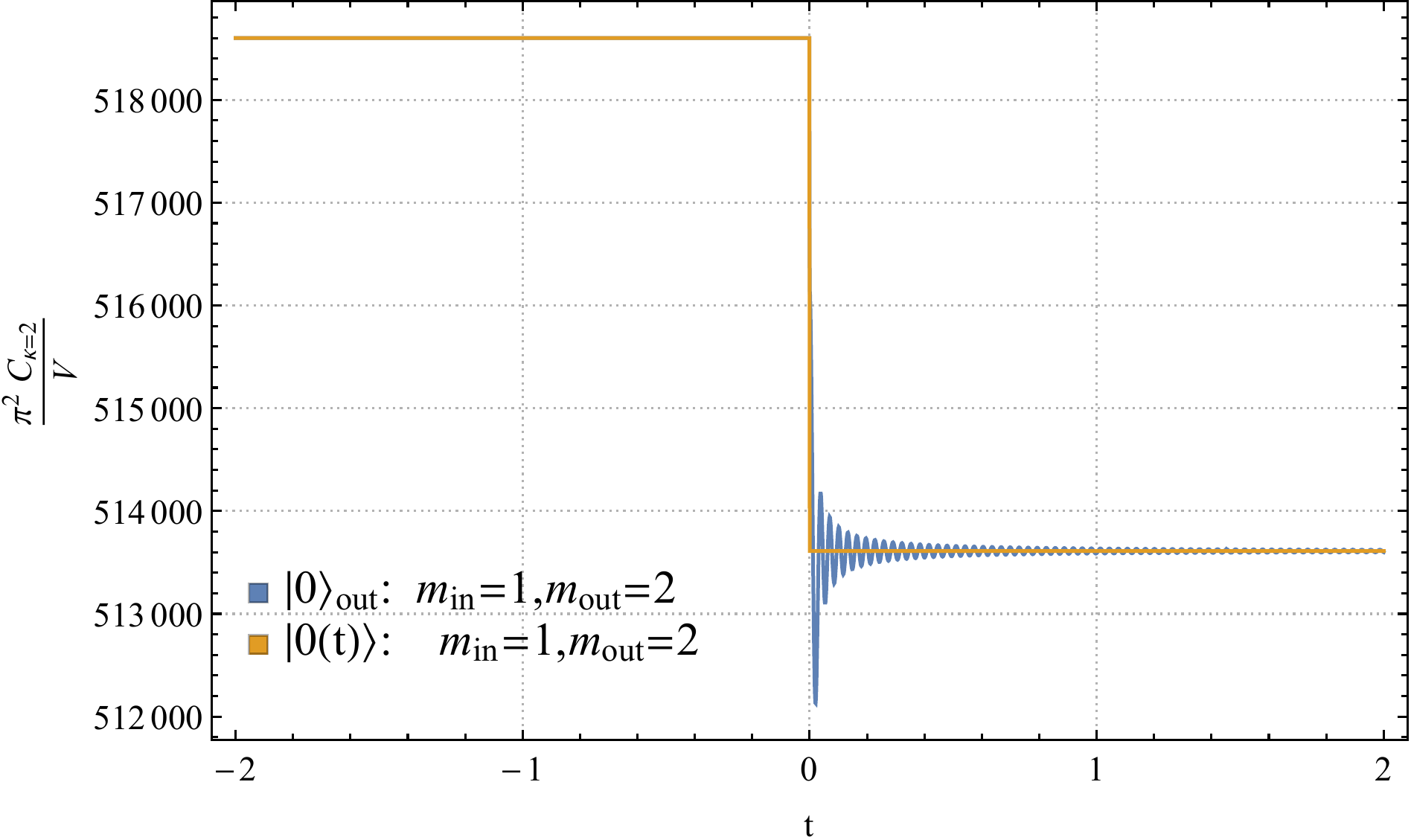}
\end{minipage}
\caption{Time evolution of the total complexity for the rotational reference state. In figure (a), we fix $m_{\text{in}}=4$ and vary $m_{\text{in}}=2,3,4,5,6$. In (b), we compare the total complexity of the incoming vacuum state with the instantaneous vacuum state.}
\label{fig2}
\end{figure}
\subsection{non-rotational invariant reference state}
Next, we vary from the reference state to a non-rotational invariant reference state, which corresponds to spinors associated with a massive state that has mass $M$ and momentum $\bm{q}$ in a given direction. Without loss of generality, we choose $\bm{q}=(0,0,q)$. We first consider the instantaneous vacuum state. According to \eq{WR} and \eq{Wvac}, the one-mode relative covariant matrix can be given by
\ba
\breve{\D}(\bm{k},t)=
\left(
\begin{array}{cccccccc}
 \frac{m M+k_z q}{\breve{\w}  E_q} & -\frac{k_x q}{\breve{\w}  E_q} & \frac{m q-M k_z}{\breve{\w}  E_q} & -\frac{M k_x}{\breve{\w}  E_q} & 0 & -\frac{k_y q}{\breve{\w}  E_q} & 0 & -\frac{M k_y}{\breve{\w}  E_q} \\
 \frac{k_x q}{\breve{\w}  E_q} & \frac{m M+k_z q}{\breve{\w}  E_q} & -\frac{M k_x}{\breve{\w}  E_q} & \frac{M k_z-m q}{\breve{\w}  E_q} & -\frac{k_y q}{\breve{\w}  E_q} & 0 & \frac{M k_y}{\breve{\w}  E_q} & 0 \\
 \frac{M k_z-m q}{\breve{\w}  E_q} & \frac{M k_x}{\breve{\w}  E_q} & \frac{m M+k_z q}{\breve{\w}  E_q} & -\frac{k_x q}{\breve{\w}  E_q} & 0 & \frac{M k_y}{\breve{\w}  E_q} & 0 & -\frac{k_y q}{\breve{\w}  E_q} \\
 \frac{M k_x}{\breve{\w}  E_q} & \frac{m q-M k_z}{\breve{\w}  E_q} & \frac{k_x q}{\breve{\w}  E_q} & \frac{m M+k_z q}{\breve{\w}  E_q} & -\frac{M k_y}{\breve{\w}  E_q} & 0 & -\frac{k_y q}{\breve{\w}  E_q} & 0 \\
 0 & \frac{k_y q}{\breve{\w}  E_q} & 0 & \frac{M k_y}{\breve{\w}  E_q} & \frac{m M+k_z q}{\breve{\w}  E_q} & -\frac{k_x q}{\breve{\w}  E_q} & \frac{m q-M k_z}{\breve{\w}  E_q} & -\frac{M k_x}{\breve{\w}  E_q} \\
 \frac{k_y q}{\breve{\w}  E_q} & 0 & -\frac{M k_y}{\breve{\w}  E_q} & 0 & \frac{k_x q}{\breve{\w}  E_q} & \frac{m M+k_z q}{\breve{\w}  E_q} & -\frac{M k_x}{\breve{\w}  E_q} & \frac{M k_z-m q}{\breve{\w}  E_q} \\
 0 & -\frac{M k_y}{\breve{\w}  E_q} & 0 & \frac{k_y q}{\breve{\w}  E_q} & \frac{M k_z-m q}{\breve{\w}  E_q} & \frac{M k_x}{\breve{\w}  E_q} & \frac{m M+k_z q}{\breve{\w}  E_q} & -\frac{k_x q}{\breve{\w}  E_q} \\
 \frac{M k_y}{\breve{\w}  E_q} & 0 & \frac{k_y q}{\breve{\w}  E_q} & 0 & \frac{M k_x}{\breve{\w}  E_q} & \frac{m q-M k_z}{\breve{\w}  E_q} & \frac{k_x q}{\breve{\w}  E_q} & \frac{m M+k_z q}{\breve{\w}  E_q} \\
\end{array}
\right)\,.
 \ea
The corresponding eigenvalues appear with a multiplicity of four and are explicitly given by
\ba
\text{spec}(\breve{\D})=\frac{(m M+p_z q)\pm i\sqrt{(p_x^2+p_y^2)(M^2+q^2)+(M p_z-m q)^2})}{E_q\breve{\w}}\,.
\ea
The corresponding one mode complexity can be expressed as
\ba
\breve{Y}(\bm{k},s,\hat{q},t)=\frac{\p}{2}-\tan^{-1}\lf(\frac{m+p_z \hat{q}}{\sqrt{(p_x^2+p_y^2)(1+\hat{q}^2)+(p_z-m \hat{q})^2}}\rt)\,,
\ea
where we set $\hat{q}=q/M$.

\begin{figure}[H]
\begin{minipage}[b]{0.5\linewidth}
\centering
\includegraphics[width=3in,height=2in]{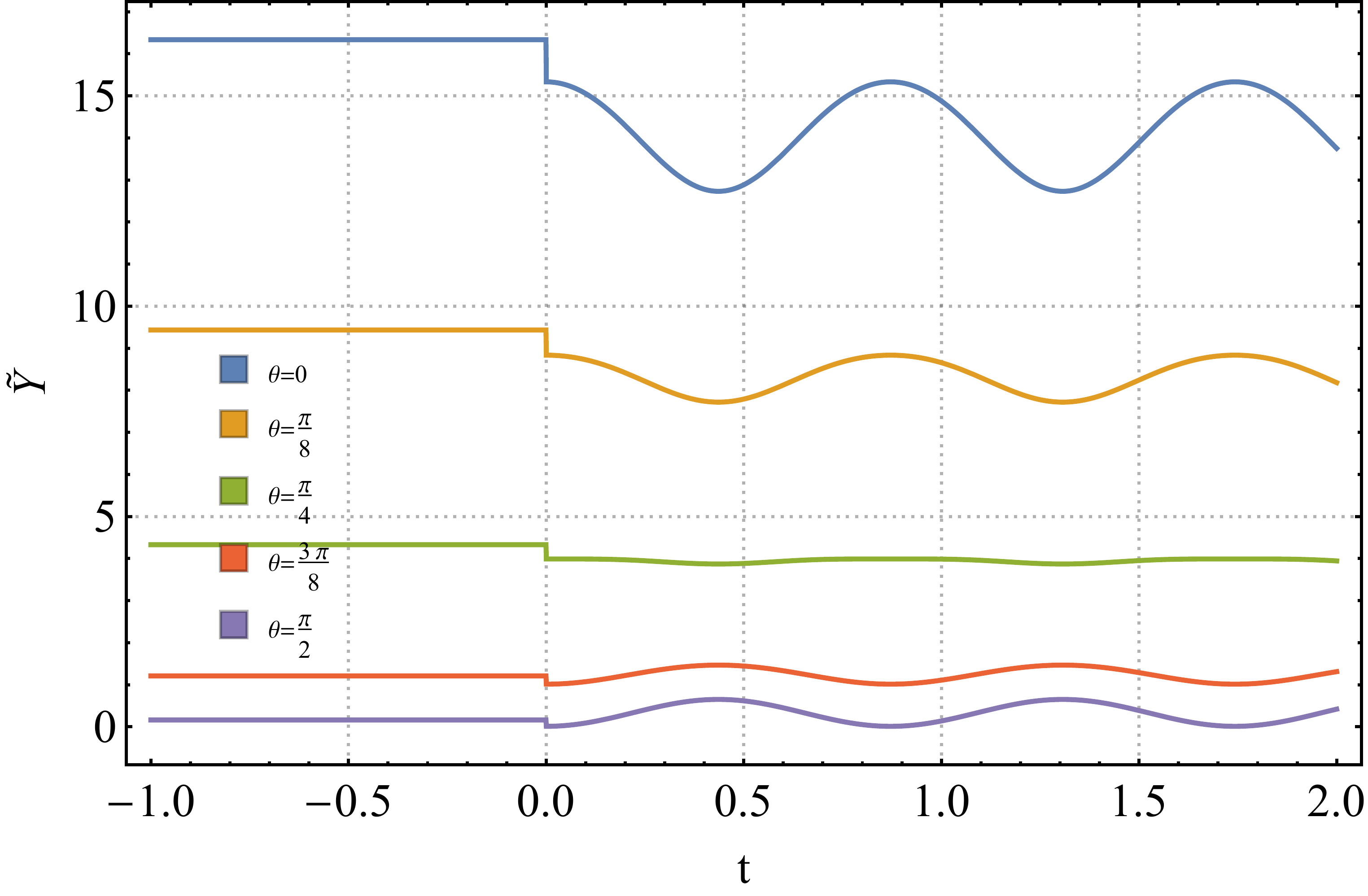}
\end{minipage}
\begin{minipage}[b]{.5\linewidth}
\includegraphics[width=3in,height=2in]{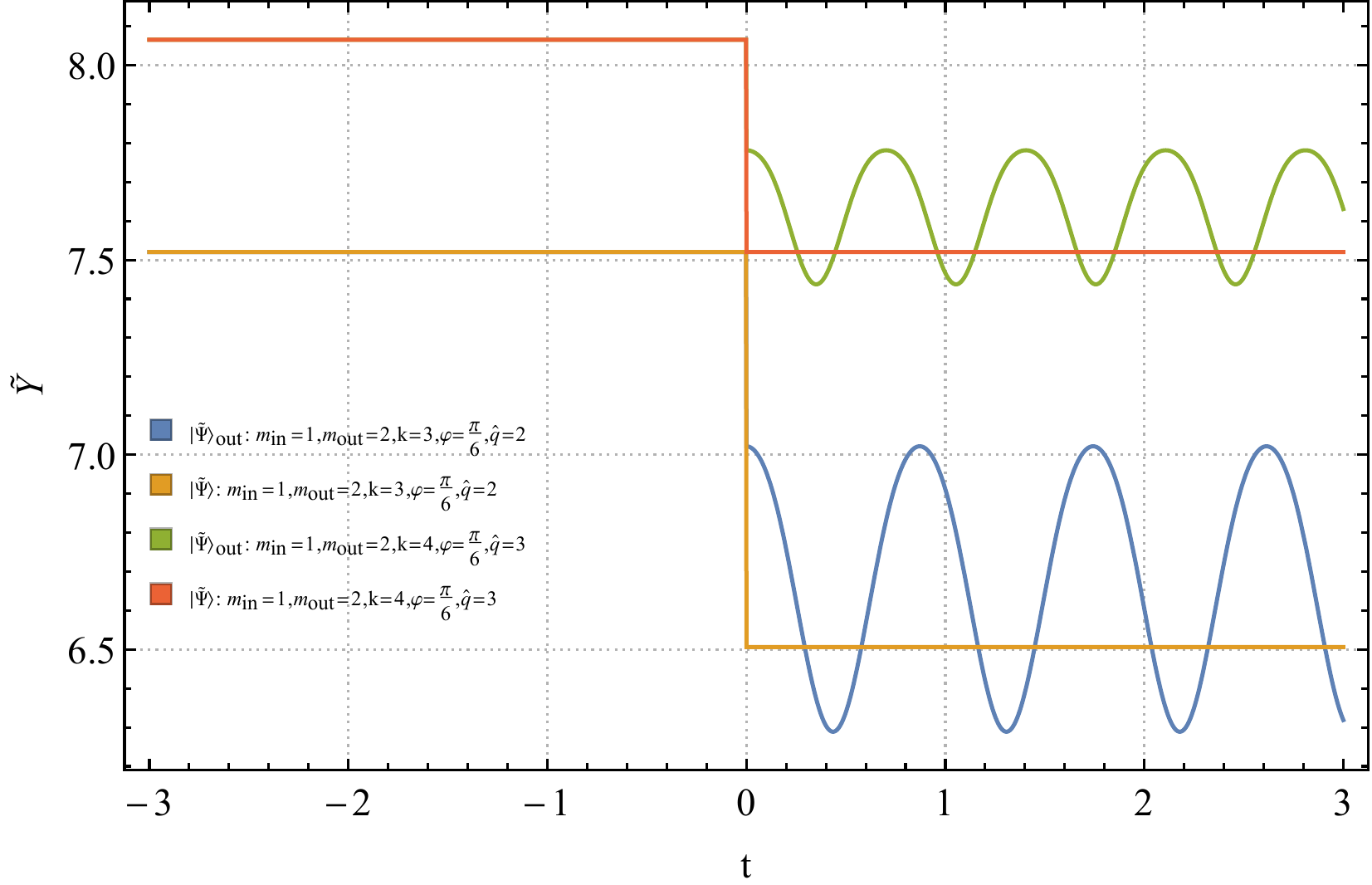}
\end{minipage}
\caption{Time evolution of the one-mode complexity for the non-rotational reference state. In figure (a), we fix $m_{\text{in}}=2, m_{\text{out}}=1, k=3, \hat{q}=1$ and vary the angle $\theta=0,\p/8,\p/4,3\p/8,\p/2$. In (b), we compare the one-mode complexity of the incoming vacuum state with the instantaneous vacuum state.}
\label{fig3}
\end{figure}

Next, we turn to the incoming  vacuum state. By \eq{WR} and \eq{WTt}, the one-mode relative covariant matrix can be written as
\ba\begin{aligned}
&\D(\bm{k})=\frac{|\f_{\bm{k}}|^2}{\w_{\text{in}}(\w_{\text{in}}+m_{\text{in}})}\times\\
&\left(\begin{smallmatrix}
\frac{p_z q \a_{\bm{k}}+M \k_{\bm{k}} }{E_q } & -\frac{p_x q \a_{\bm{k}}+M p_y \d_{\bm{k}} }{E_q } & \frac{q \k_{\bm{k}} -M p_z \a_{\bm{k}}}{E_q } & \frac{p_y q \d_{\bm{k}} -M p_x \a_{\bm{k}}}{E_q } & \frac{M p_z \d_{\bm{k}} }{E_q } & \frac{M p_x \d_{\bm{k}} -p_y q \a_{\bm{k}}}{E_q } & \frac{p_z q \d_{\bm{k}} }{E_q } & -\frac{M p_y \a_{\bm{k}}+p_x q \d_{\bm{k}} }{E_q } \\
 \frac{p_x q \a_{\bm{k}}+M p_y \d_{\bm{k}} }{E_q } & \frac{p_z q \a_{\bm{k}}+M \k_{\bm{k}} }{E_q } & \frac{p_y q \d_{\bm{k}} -M p_x \a_{\bm{k}}}{E_q } & \frac{M p_z \a_{\bm{k}}-q \k_{\bm{k}} }{E_q } & \frac{M p_x \d_{\bm{k}} -p_y q \a_{\bm{k}}}{E_q } & -\frac{M p_z \d_{\bm{k}} }{E_q } & \frac{M p_y \a_{\bm{k}}+p_x q \d_{\bm{k}} }{E_q } & \frac{p_z q \d_{\bm{k}} }{E_q } \\
 \frac{M p_z \a_{\bm{k}}-q \k_{\bm{k}} }{E_q } & \frac{M p_x \a_{\bm{k}}+p_y q \d_{\bm{k}} }{E_q } & \frac{p_z q \a_{\bm{k}}+M \k_{\bm{k}} }{E_q } & \frac{M p_y \d_{\bm{k}} -p_x q \a_{\bm{k}}}{E_q } & \frac{p_z q \d_{\bm{k}} }{E_q } & \frac{M p_y \a_{\bm{k}}-p_x q \d_{\bm{k}} }{E_q } & -\frac{M p_z \d_{\bm{k}} }{E_q } & -\frac{p_y q \a_{\bm{k}}+M p_x \d_{\bm{k}} }{E_q } \\
 \frac{M p_x \a_{\bm{k}}+p_y q \d_{\bm{k}} }{E_q } & \frac{q \k_{\bm{k}} -M p_z \a_{\bm{k}}}{E_q } & \frac{p_x q \a_{\bm{k}}-M p_y \d_{\bm{k}} }{E_q } & \frac{p_z q \a_{\bm{k}}+M \k_{\bm{k}} }{E_q } & \frac{p_x q \d_{\bm{k}} -M p_y \a_{\bm{k}}}{E_q } & \frac{p_z q \d_{\bm{k}} }{E_q } & -\frac{p_y q \a_{\bm{k}}+M p_x \d_{\bm{k}} }{E_q } & \frac{M p_z \d_{\bm{k}} }{E_q } \\
 -\frac{M p_z \d_{\bm{k}} }{E_q } & \frac{p_y q \a_{\bm{k}}-M p_x \d_{\bm{k}} }{E_q } & -\frac{p_z q \d_{\bm{k}} }{E_q } & \frac{M p_y \a_{\bm{k}}+p_x q \d_{\bm{k}} }{E_q } & \frac{p_z q \a_{\bm{k}}+M \k_{\bm{k}} }{E_q } & -\frac{p_x q \a_{\bm{k}}+M p_y \d_{\bm{k}} }{E_q } & \frac{q \k_{\bm{k}} -M p_z \a_{\bm{k}}}{E_q } & \frac{p_y q \d_{\bm{k}} -M p_x \a_{\bm{k}}}{E_q } \\
 \frac{p_y q \a_{\bm{k}}-M p_x \d_{\bm{k}} }{E_q } & \frac{M p_z \d_{\bm{k}} }{E_q } & -\frac{M p_y \a_{\bm{k}}+p_x q \d_{\bm{k}} }{E_q } & -\frac{p_z q \d_{\bm{k}} }{E_q } & \frac{p_x q \a_{\bm{k}}+M p_y \d_{\bm{k}} }{E_q } & \frac{p_z q \a_{\bm{k}}+M \k_{\bm{k}} }{E_q } & \frac{p_y q \d_{\bm{k}} -M p_x \a_{\bm{k}}}{E_q } & \frac{M p_z \a_{\bm{k}}-q \k_{\bm{k}} }{E_q } \\
 -\frac{p_z q \d_{\bm{k}} }{E_q } & \frac{p_x q \d_{\bm{k}} -M p_y \a_{\bm{k}}}{E_q } & \frac{M p_z \d_{\bm{k}} }{E_q } & \frac{p_y q \a_{\bm{k}}+M p_x \d_{\bm{k}} }{E_q } & \frac{M p_z \a_{\bm{k}}-q \k_{\bm{k}} }{E_q } & \frac{M p_x \a_{\bm{k}}+p_y q \d_{\bm{k}} }{E_q } & \frac{p_z q \a_{\bm{k}}+M \k_{\bm{k}} }{E_q } & \frac{M p_y \d_{\bm{k}} -p_x q \a_{\bm{k}}}{E_q } \\
 \frac{M p_y \a_{\bm{k}}-p_x q \d_{\bm{k}} }{E_q } & -\frac{p_z q \d_{\bm{k}} }{E_q } & \frac{p_y q \a_{\bm{k}}+M p_x \d_{\bm{k}} }{E_q } & -\frac{M p_z \d_{\bm{k}} }{E_q } & \frac{M p_x \a_{\bm{k}}+p_y q \d_{\bm{k}} }{E_q } & \frac{q \k_{\bm{k}} -M p_z \a_{\bm{k}}}{E_q } & \frac{p_x q \a_{\bm{k}}-M p_y \d_{\bm{k}} }{E_q } & \frac{p_z q \a_{\bm{k}}+M \k_{\bm{k}} }{E_q } \\
\end{smallmatrix}\right)\,.\nn
\end{aligned}\ea
Note that this relative covariant matrix only depends on the quantity $\hat{q}$. Without loss of generality, next we set $M=1$. For simplicity, we consider the one-mode complexity of per momenta $\tilde{Y}(\bm{k},t)$, such that
\ba
\math{C}_{\k=2}=\frac{V}{4}\int^{\L}\frac{d^3k}{(2\p)^3} \tilde{Y}(\bm{k},t)\,.
\ea
By the numerical analysis, we show some relevant results in \fig{fig3}. It's easy to show that the one-mode complexity will share similar behaviors with the rotational reference state under variation of the parameters $\bm{k}, m_\text{in}$ and $m_\text{out}$. Moreover, by \fig{fig3},  one can find that the one-mode complexity also depends on the angle $\theta$ between the momenta $\bm{k}$ and $\bm{q}$. With the growth of the angle, the amplitude as well as the jump value decreases, but this variation will not affect the sign of the jump value. However, from \fig{fig2}, one can find that there exists a turning point of $\hat{q}$ which shifts the sign of the jump value from positive to negative, even though we fix the sign of $m_-$. That is to say, for the non-rotational reference state, by choosing the value of $\hat{q}$, we can change the relationship between the jump value and the mass difference, which can not be realized in the rotational case.

Moreover, as shown in (c), at $t>0$, the equilibrium position does not locate on the position of the instantaneous vacuum state,  which means that the total complexity will not saturate the instantaneous result at the late time. This is actually different from the rotational invariant case in the last section.

\begin{figure}[H]
\begin{minipage}[b]{0.5\linewidth}
\centering
\includegraphics[width=3in,height=2in]{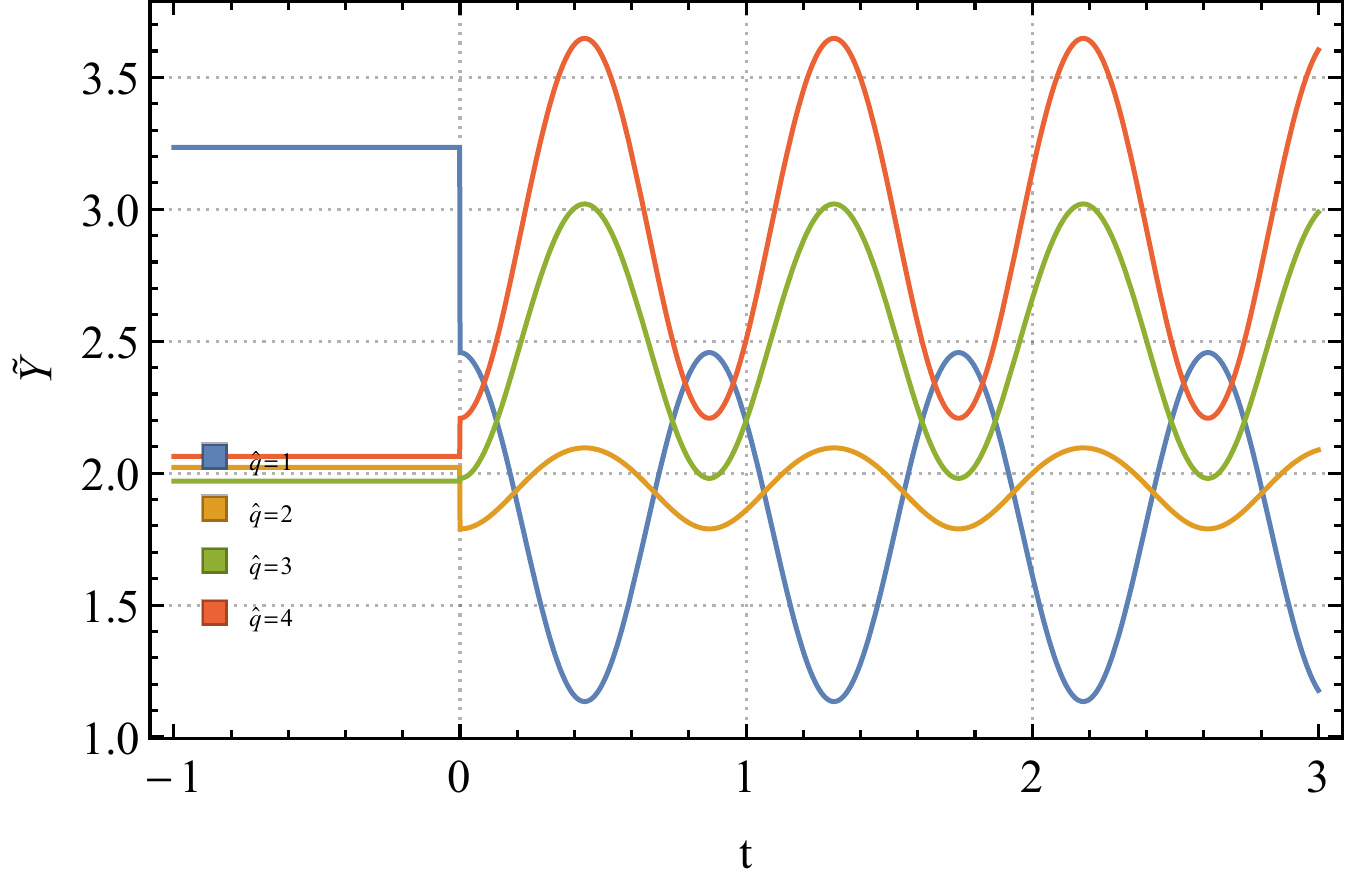}
\end{minipage}
\begin{minipage}[b]{.5\linewidth}
\includegraphics[width=3in,height=2in]{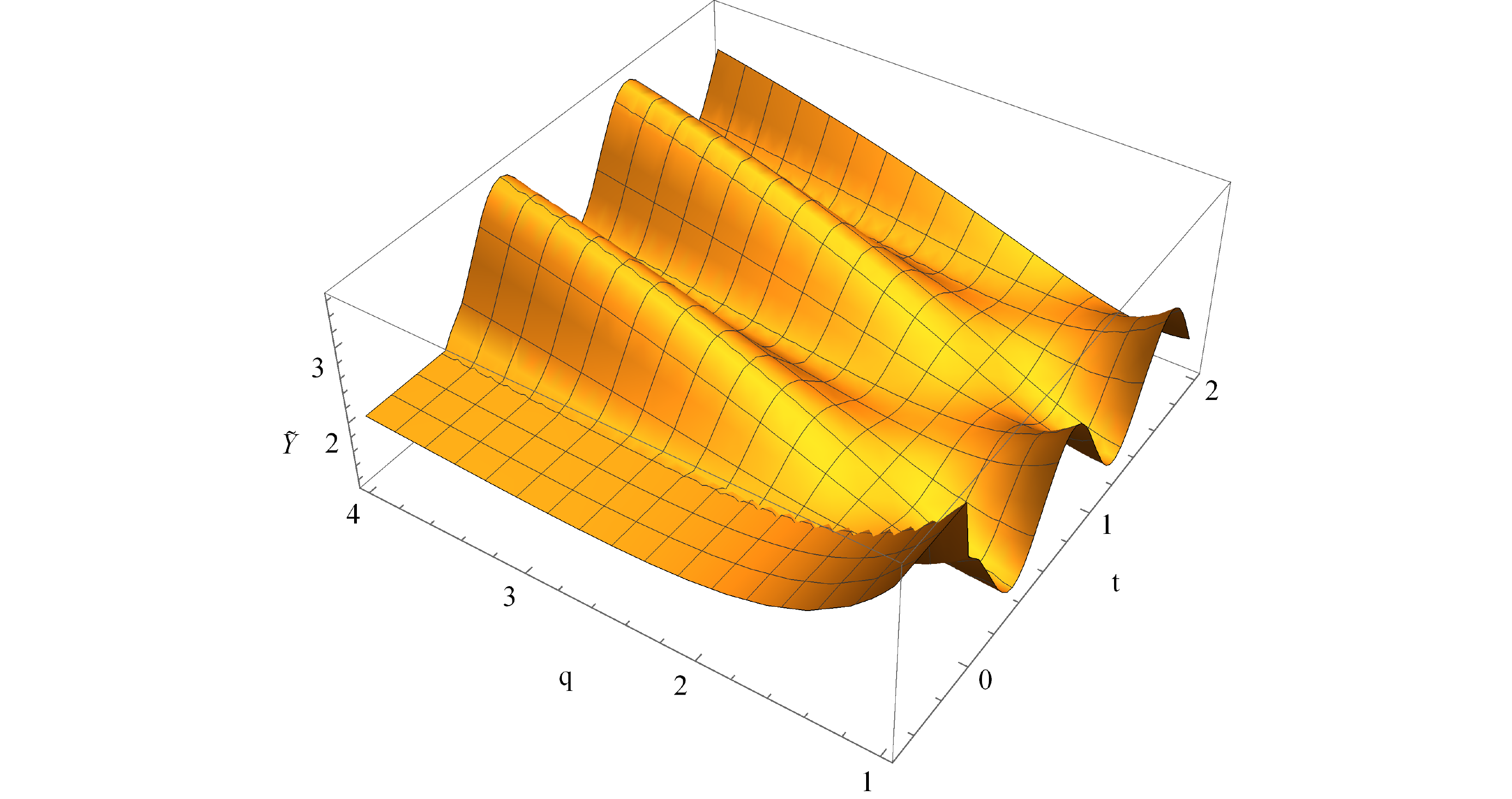}
\end{minipage}
\caption{Time evolution of the one-mode complexity from the non-rotational reference state to the incoming vacuum state, where we fix $m_{\text{in}}=1, m_{\text{out}}=2, k=3, \theta=\p/3$ and vary the momenta of the reference state $\hat{q}=1,2,3,4$}
\label{fig2}
\end{figure}

\subsection{Excited states as target state}
In this section, we consider the target state
 $\bra{\Y}=a_{\bm{k}}^{s\dag}b_{-\bm{k}}^{s\dag}\bra{0}$ that is stated in section.\ref{eset}. By \eq{WR} and \eq{WY}, one can obtain the relative covariant matrix
\ba\begin{aligned}
\D_{\Y}=\frac{|\f_{\bm{k}}|^2}{\w_{\text{in}}(\w_{\text{in}}+m_{\text{in}})}\left(
\begin{array}{cccccccc}
-\g_{\bm{k}}  & k_xk_z&k_z\a_{\bm{k}}  & 0 & -k_z \d_{\bm{k}}  & k_yk_z& 0 & 0 \\
 k_xk_z& \g_{\bm{k}}  & 0 &k_z\a_{\bm{k}}  & -k_yk_z& -k_z \d_{\bm{k}}  & 0 & 0 \\
 -k_z \a_{\bm{k}}  & 0 & -\g_{\bm{k}}  & k_xk_z& 0 & 0 &k_z\d_{\bm{k}}  & k_yk_z\\
 0 & -k_z \a_{\bm{k}}  & k_xk_z& \g_{\bm{k}}  & 0 & 0 & -k_yk_z&k_z\d_{\bm{k}}  \\
k_z\d_{\bm{k}}  & -k_yk_z& 0 & 0 & -\g_{\bm{k}}  & k_xk_z&k_z\a_{\bm{k}}  & 0 \\
 k_yk_z&k_z\d_{\bm{k}}  & 0 & 0 & k_xk_z& \g_{\bm{k}}  & 0 &k_z\a_{\bm{k}}  \\
 0 & 0 & -k_z \d_{\bm{k}}  & -k_yk_z& -k_z \a_{\bm{k}}  & 0 & -\g_{\bm{k}}  & k_xk_z\\
 0 & 0 & k_yk_z& -k_z \d_{\bm{k}}  & 0 & -k_z \a_{\bm{k}}  & k_xk_z& \g_{\bm{k}}  \\
\end{array}
\right)\,,
\end{aligned}\ea
The corresponding eigenvalues appear in two quadruples $\lf(e^{2i\q_1},e^{2i\q_2},e^{-2i\q_1},e^{-2i\q_2}\rt)$  are given by
\ba
e^{\pm 2i\q_1}&=&\frac{|\f_{\bm{k}}|^2}{\w_{\text{in}}(\w_{\text{in}}+m_{\text{in}})}\lf(\sqrt{k^4-k^2(k_z^2-2\k_{\bm{k}})+\k_{\bm{k}}(\k_{\bm{k}}-2k_z^2)}\pm i |k_z| \sqrt{\a_{\bm{k}}^2+\d_{\bm{k}}^2}\rt)\,,\\
e^{\pm 2i\q_2}&=&-\frac{|\f_{\bm{k}}|^2}{\w_{\text{in}}(\w_{\text{in}}+m_{\text{in}})}\lf(\sqrt{k^4-k^2(k_z^2-2\k_{\bm{k}})+\k_{\bm{k}}(\k_{\bm{k}}-2k_z^2)}\pm i |k_z| \sqrt{\a_{\bm{k}}^2+\d_{\bm{k}}^2}\rt)\,.
\ea
Similarly, the one mode contribution to the complexity of each spin can be given by
\ba
2\q_1&=&\tan^{-1}\lf(\frac{|k_z| \sqrt{\a_{\bm{k}}^2+\d_{\bm{k}}^2}}{\sqrt{k^4-k^2(k_z^2-2\k_{\bm{k}})+\k_{\bm{k}}(\k_{\bm{k}}-2k_z^2)}}\rt)\,,\\
2\q_2&=&\p-\tan^{-1}\lf(\frac{|k_z| \sqrt{\a_{\bm{k}}^2+\d_{\bm{k}}^2}}{\sqrt{k^4-k^2(k_z^2-2\k_{\bm{k}})+\k_{\bm{k}}(\k_{\bm{k}}-2k_z^2)}}\rt)\,.
\ea
Then, the contribution from this mode can be written as
\ba
\tilde{Y}(\bm{k},t)^2=(2\q_1)^2+(2\q_2)^2\,,
\ea
Consider the total complexity, this excited state mode only makes a finite perturbation to the vacuum complexity. Thus, we can only consider the difference between the complexity of the excited state and that of the vacuum state, $i,e.$,
\ba\begin{aligned}
\D\math{C}_{\k=2}(t)&=\math{C}_{\k=2}(t)-\hat{\math{C}}_{\k=2}(t)\\
&=\tilde{Y}(\bm{k},t)^2-2Y(\bm{k},s,t)^2.
\end{aligned}\ea
As illustrated in (a) and (b), one can find that this complexity goes up monotonically with the angle $\theta$ as well as the momenta $k$, but it actually doesn't affect the sign of the jump value of the complexity, which means that the sign of the jump value only depends on the sign of $m_-$. But interestingly, according to (c), one can find that except for the critical value $m_-=0$, there exists another critical value where the wave crest will suddenly occur in the trough of the wave.
\begin{figure}[H]
\centering
\includegraphics[width=6.5in,height=4in]{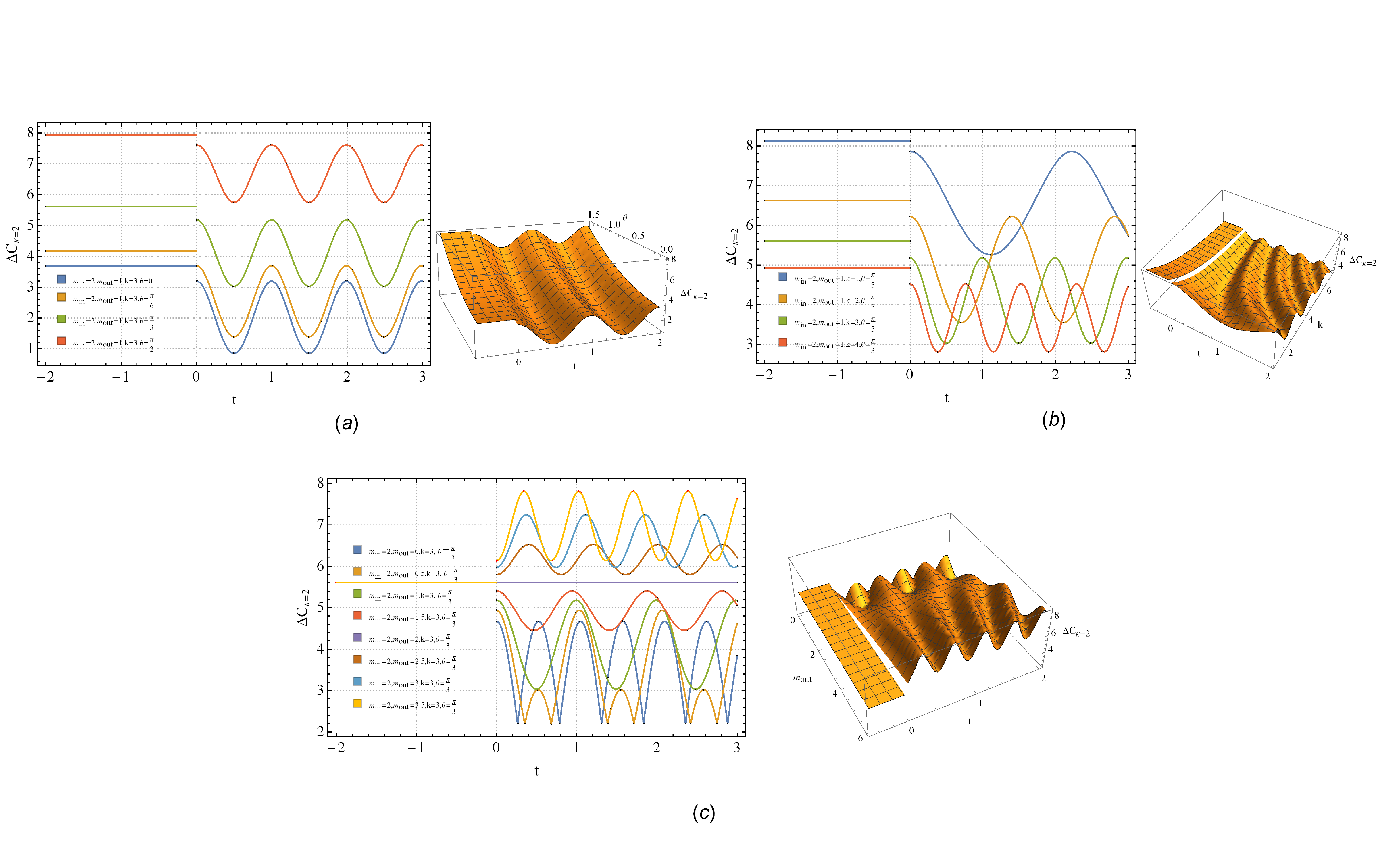}
\caption{Time evolution of the complexity from the rotational reference state to the excited state $\bra{\Y}$. In figure (a), $m_{\text{in}}=2, m_{\text{out}}=1, k=3$ is fixed and $\theta$ is changed as shown in the figure. In (b), we fix $m_{\text{in}}=2, m_{\text{out}}=1, \theta=\p/3$ and vary $k$ as shown in the figure. In (c), we fix $m_{\text{in}}=2, k=3, \theta=\p/3 $ and vary $m_{\text{out}}=0,0.5,1,1.5,2,2.5,3,3.5$.}
\label{figf}
\end{figure}
\section{Conclusions}\label{conclusions}
In this paper, we have investigated the time evolution of the circuit complexity in a Fermion system with a mass quench. It has been pointed in the introduction that this model can be regarded as a toy model for the study of the complexity of a thermodynamic system. Before computing the complexity of these states, we first review the counting method which is given by Hackl $et\, al.$, and demonstrate that this result can be adapted to all of the compact transformation group $G$. Then, we apply this result to evaluating the time evolution of the complexity of some particular vacuum states. We show that, for the rotational reference state, the total complexity of the incoming vacuum state will saturate the value of the instantaneous vacuum state at the late time, with a typical timescale to achieve the final stable state. Moreover, we find that the jump value under the sudden quench is directly proportional to the mass difference $\d m$.  Note that the incoming vacuum state corresponds to the AdS vacuum, and the late time thermal state corresponds to the AdS-Vaidya hole. To connect our result to the holograph system, we propose that the circuit complexity of a free system is dual to the complexity growth rate of a strongly coupled system \eq{fs}. Under this conjecture, our result shares a similar behaviour with the holograph complexity growth rate in an AdS-Vaidya black hole equipped with a shock wave\cite{CMM}. Furthermore, we illustrate that apart from the divergence contributed by \eq{divC1}, which shares a similar formalism with the static Dirac vacuum state\cite{C}, there also exist some divergent parts contributed by the amplitude. Then, we evaluate the complexity of the incoming vacuum state for a non-rotational invariant reference state. Unlike the case of rotational reference state, we can change the relationship between the jump value and the mass difference in the non-rotational case. Moreover, the total complexity will not saturate the instantaneous result at the late time. Finally, we compute the complexity from a rotational reference state to an incoming exited state and then find that there exists a critical value of $m_-$ in which a wave crest will suddenly occur in the trough of the wave.

\section*{Acknowledgments}
 This research was supported by NSFC Grants No. 11775022 and 11375026.

\end{document}